\documentclass[12pt,preprint]{aastex}

\slugcomment{To appear in ApJ, December 1, 2003}

\shorttitle{Radiative Transfer in Protostars: Evolution}
\shortauthors{Whitney, Wood, Bjorkman, \& Cohen}

\begin{document}


\title{
2-D Radiative Transfer in Protostellar Envelopes:  II.  An Evolutionary
Sequence 
}


\author{Barbara A. Whitney\altaffilmark{1}, 
Kenneth Wood\altaffilmark{2}, J. E. Bjorkman\altaffilmark{3},
\& Martin Cohen\altaffilmark{4}} 

\altaffiltext{1}{Space Science Institute, 3100 Marine Street, Suite A353, 
Boulder, CO~80303; bwhitney@colorado.edu}

\altaffiltext{2}{School of Physics \& Astronomy, University of St Andrews, 
North Haugh, St Andrews, Fife, KY16 9AD, Scotland the Brave; 
kw25@st-andrews.ac.uk}

\altaffiltext{3}{Ritter Observatory, MS 113, Department of Physics \& Astronomy, 
University of Toledo, Toledo, OH 43606-3390; jon@physics.utoledo.edu}

\altaffiltext{4}{Radio Astronomy Laboratory, 601 Campbell Hall, University of California,
Berkeley, CA 94720; mcohen@astro.berkeley.edu
}

\begin{abstract}

We present model spectral energy distributions, colors, polarization, and images 
for an evolutionary sequence of a low-mass
protostar from the early collapse stage (Class 0) to the remnant disk stage 
(Class III).
We find a substantial overlap in colors and SEDs between protostars embedded in envelopes
(Class 0-I) and T Tauri disks (Class II), especially at mid-IR wavelengths.  
Edge-on Class I-II sources show double-peaked spectral energy distributions, with a short-wavelength
hump due to scattered light and the long-wavelength hump due to thermal
emission.  These are the bluest sources in mid-IR color-color diagrams.

Since Class 0 and I sources are diffuse, the size of the aperture over which fluxes are integrated
has a substantial effect on the computed colors, with larger aperture results showing
significantly bluer colors.  Viewed through large apertures, the Class 0 colors
fall in the same regions of mid-IR color-color diagrams as Class I sources,
and are even bluer than Class II-III sources in some colors.   
It is important to take this into
account when comparing color-color diagrams of star formation regions at
different distances, or different sets of observations of the same region.  
However the near-IR polarization of the Class 0 sources is much higher than
the Class I-II sources, providing  a means to separate these evolutionary
states.
 

We varied the grain properties in the circumstellar envelope, allowing for
larger grains in the disk midplane and smaller in the envelope.  In comparing to
models with the same grain properties throughout we find that the SED of the Class 0
source is sensitive to the grain properties of the envelope only--that is, grain growth in the disk in Class 0 sources cannot be detected from the SED.  Grain growth in disks of 
Class I sources can be detected at wavelengths greater than 100 $\mu$m.

Our image calculations predict that the diffuse emission from edge-on Class I and II 
sources should be detectable in the mid-IR with the Space Infrared Telescope Facility
(SIRTF) in nearby star forming regions 
(out to several hundred parsecs).

\end{abstract}


\keywords{radiative transfer---stars: formation---stars: imaging---
stars: pre-main sequence---circumstellar matter---dust, extinction---
polarization}


\section{Introduction}

Low-mass protostars have been grouped into an evolutionary sequence,
Class 0 through Class III, based on their spectral energy distributions 
(Lada \& Wilking 1984; Lada 1987;  Myers et al. 1987; Strom et al. 1989;
Adams, Lada, \& Shu 1987; Andr\'e, Ward-Thompson, \& Barsony 1993).
Class 0 sources are in the early protostellar collapse stage, with 
spectral energy distributions (SEDs)
that resemble blackbodies with T $\le$ 30K (Andr\'e et al. 1993).
The majority of the source mass resides in the infalling
envelope.  Even at this early stage of collapse  (of order a few $\times$ 10$^4$ yr), 
these sources exhibit powerful, bipolar molecular flows.
The Class I stage is a later stage of protostellar collapse
which lasts a few $\times$ 10$^5$ yrs.  These
sources display very broad SEDs that peak near 100 $\mu$m. 
Their envelope masses are similar to the mass of the central pre-main-sequence
core.  They have well-developed accretion disks and their envelopes
have bipolar cavities excavated
by outflows (Terebey, Chandler, \& Andr\'e 1993, Kenyon et al. 1993b, Tamura et al. 1991).  
The Class II stage is characterized by
the presence of excess infrared emission above that expected for a stellar
photosphere, with the SED emission peak occuring in the near-infrared.
This is the signature of a Classical T Tauri star surrounded by an
accretion disk.  
Infall from the cloud has ceased due to dispersal of the remnant infall envelope
by the combined effects of infall and outflow.
Class II sources have ages of about $10^6 - 10^7$ yrs (Strom et al. 1989).
The Class III sources, also known as ``weak-lined'' T Tauri stars,
have SED's that resemble a stellar photosphere.  Mid-IR radiation
from remant disk material, if it exists, should be detectable by the upcoming Space
Infrared Telescope Facility (SIRTF).  
It is not known for certain
whether Class III sources are more evolved than the Class II sources,
or whether they have simply lost most of their circumstellar material
on a faster timescale.
The Class III stage can
last for about $10^7$ years, ending in a Zero-Age Main-Sequence star (Palla 1999).

Interpretations of the SEDs of pre-main sequence stars have traditionally relied on 1-D or 1.5-D radiative
transfer models of spherical envelopes for Class 0-I sources (Adams et al. 1987,
Kenyon, Calvet, \& Hartmann 1993a; Jayawardhana, Hartmann, \& Calvet 2001) and
flat or flared-disk models for Class II sources 
(Kenyon \& Hartmann 1987;
 Chiang \& Goldreich 1997; D'Alessio 
et al. 1998, 1999; Dullemond, Dominik, \& Natta 2001; Dullemond 2002;
Dullemond \& Natta 2003).   
While the Class II models do a good job matching observations of pure
disk sources, the models for younger sources generally
do not take
into account bipolar cavities created by outflows, vertically extended disks in Class 0-I
sources, inclination
effects, or intermediate
stages between Class I (envelopes) and II (disks) sources.
These models appeared
sufficient to explain the previously sparsely sampled SEDs, though it was clear early
on that they failed to account for the large optical/near-IR flux
from most Class I sources (Adams et al. 1987).  Kenyon et al. (1993b) showed that bipolar
cavities allow a sufficent amount of near-IR flux to scatter out to match the observations.
Now with more complete SED information on several sources showing
more complicated SEDs (Liu et al. 1996, Chandler, Barsony, \& Moore 1998,
Rebull et al. 2003, Wolf, Padgett, \& Stapelfeldt 2003), and with the 
upcoming SIRTF mission which
will sample more frequencies in the mid-IR,
it is clear that 2-D and 3-D models
are required to interpret these sources.

In a previous paper (Whitney et al. 2003, hereafter Paper I), 
we showed that including a flared disk and bipolar cavity in models
of Class I sources
leads to more complicated SEDs than those from 1-D models or 2-D models with rotationally-
flattened envelopes only (Efstathiou \& Rowan-Robinson 1991).
We showed that the near-IR and mid-IR colors of 2-D models are 
substantially bluer than
those from 1-D models for a given envelope mass.  
Thus, using 1-D models to interpret sources that
are not spherical will lead to an underestimate of the mass of the
envelope and an overestimate of the evolutionary state.  

This paper follows up Paper I with models comprising an evolutionary sequence
from Class 0 - III.  Our goal is to provide more accurate interpretations
of observations.  
We find substantial overlap between colors of sources at different evolutionary stages in the mid-IR wavelengths
that SIRTF will observe.
The information provided by SEDs alone may not be sufficient to interpret a source's 
evolutionary state, since, for example,
a more pole-on Class I source can resemble an edge-on Class II source.
Combining
more information from images and polarization, which help determine
inclination and size, can resolve this degeneracy. 
\S2 of this paper describes the models, \S 3 describes the resulting
SEDs, colors, images and polarization spectra, and \S 4 summarizes
the results.  

\section{Models}

\subsection{Circumstellar geometry}

For the infalling envelopes, we use the density structure of a 
rotating envelope in free-fall collapse (Ulrich 1976; equation 1 of Paper I).  
The disk density is a flared disk in hydrostatic equilibrium (Shakura \& 
Sunyaev
1973, Pringle 1981, equation 3 of Paper I).
The envelopes have curved bipolar cavities filled with constant density gas 
and dust.   The cavity shape follows $z = a \varpi^b$, 
where $\varpi=\sqrt{x^2 + y^2}$ is the cylindrical radius.  

We computed six models comprising an evolutionary sequence
from Class 0 to III. 
This evolutionary sequence is characterized by a decreasing envelope infall rate,
increasing disk radius, increasing bipolar cavity opening angle, and decreasing
cavity density. 
The density structures for each model are shown in the right-hand panels of Figure 3.  For more detailed
plots of the disk/envelope density in the inner regions, see
Paper I, figure 2.
Table 1 shows the parameters common to all models.
The stellar parameters are typical of a low-mass T Tauri star.
The disk parameters are based on models of T Tauri disks in hydrostatic
equilibrium (D'Alessio et al. 1998) and models of observed images 
(Cotera et al.
2001, Wood et al. 2002a).  
The disk density is proportional to $\varpi^{-\alpha}$.  The disk scale height increases
with radius as  $\varpi^{\beta}$ scaled to the value of $h_0$ at the stellar radius,
given in Table 1.
The disk viscosity parameter, $\alpha_{disk}$, is taken from models of average
passive T Tauri
accretion disks (D'Alessio et al. 1999), based on the $\alpha$-disk prescription 
(Shakura \& Sunyaev 1973,
Kenyon \& Hartmann 1987, Bjorkman 1997; Hartmann 1998; see equations 5-6 of Paper I).  
This is used in our calculation of disk
accretion rate and luminosity (Table 2).  

Table 2 shows the parameters that vary between the six models.
The envelope infall rates are based on previous models of Class 0 and I sources 
(Adams et al. 1987, Kenyon et al. 1993a,b, Jayawardhana et al. 2001,
Stark et al. 2003).
The envelope mass is determined by a combination of several parameters: envelope infall
rate, centrifugal radius (= disk outer radius), outer envelope radius, cavity
size and shape and density.  In the models with infalling envelopes (Class 0-I), the envelope mass grows with radius as $r^{1.5}$.
Thus, the envelope mass tabulated is only that within 5000 AU, our chosen outer envelope
radius.
The envelope inner radius and disk inner radius are chosen to be at or slightly larger
than the dust destruction radius (except for the Class III source);
this is larger in the more dense disks.
The Class III disk inner radius is chosen to be 50 stellar radii since evolved
disks can have large inner holes (e.g., HR 4796, Koerner et al. 1998)
The disk mass is chosen to be 0.01 $M_\sun$ for all evolutionary states except Class III.
This mass is typical for Class 0-II sources (Beckwith et al. 1990, Terebey et al. 1993,
D'Alessio, Calvet, \& Hartmann 2001, Looney, Mundy, \&
Welch 2003).
The disk size is expected to grow with time for envelopes which initially rotated 
as a solid-body.
Our choices of disk size are based on previous modeling and observations (Beckwith et al. 1990, Kenyon et al. 1993a,b,
Stark et al. 2003, Padgett et al. 1999).
The envelope centrifugal radius is chosen to be the same as the disk outer
radius.
The disk accretion rate and contribution to luminosity varies between the models due
to different disk densities.  
We note that its effect on the SEDs and images is small.
The Class III disk mass is chosen to be $10^{-6} M_\sun$.  For more complete treatments
of Class III disks and debris disks exploring a wider range of parameter space
in mass and dust properties, see Wood et al. (2002b) and Wolf \& Hillenbrand (2003).
The bipolar cavity densities are based on observations of molecular outflow densities
(Moriarty-Schieven et al. 1992, 1995a,b) and our assumption that these would decrease 
with evolution.
The cavity opening angles are also expected to increase with age and our choices
are based on images of molecular outflows and envelopes (Chandler et al. 1996,
Tamura et al. 1996, Hogerheijde et al. 1998, Padgett et al. 1999) and
models of images (Stark et al. 2003).
The 90 degree opening angle for the Class II and III sources is used to give
a constant density envelope at typical molecular cloud ambient densities.  The density
of the Class III ambient cloud is lower than the other sources since the disk density
is low and it would not be surprising to see Class III sources in less dense
regions of the molecular cloud.

\subsection{Radiative Transfer}

The radiative transfer method is the same as described in Paper I.  It uses the radiative 
equilibrium solution method of Bjorkman \& Wood (2001) incorporated into a 
3-D spherical-polar grid.  
Complicated 3-D density distributions can be solved with the same ease as 
1-D spherical densities.  
The models properly account for scattering and inclination effects.
Most of the source luminosity comes from the central star, with a small amount of
accretion luminosity from the disk.  The input stellar spectrum is a Kurucz model atmosphere
(Kurucz 1994)
for the stellar parameters given in Table 1.
Both the stellar and disk accretion luminosity are reprocessed by the dusty disk and envelope.
An addition to the models in this paper is the capability to 
include different grains
in different regions.  A grain model is assigned to each grid cell.  Several variables
used for the temperature calculation in a cell, e.g., the Rosseland mean opacity, are computed
for each grain model used in the calculation.  

\subsection{Grain Properties}

Previous modeling of SEDs of disks has shown that to fit images and SEDs
of Class II sources (disks) requires large grains with maximum sizes up to 1 mm 
(D'Alessio et al. 2001, Wood et al. 2002a, Wolf et al. 2003).  
On the other hand, to match
the colors, images and polarization maps of Class I
sources (mainly envelopes) in the Taurus molecular cloud, grain models similar to the 
diffuse ISM work
reasonably well (Kenyon et al. 1993b, Whitney et al. 1997,
Lucas \& Roche 1997, 1998, Stark et al. 2003, Wolf et
al. 2003).  For this reason we allow the grain properties to vary in different 
regions, as shown in Figure 1.  
For the dense regions of the disk (n$_{H2}> 10^8$ cm$^{-3}$), we use the large 
grain
model that Wood et al. (2002a) used to fit the HH30 disk SED.   For the less dense regions
of the disk, we use the same grain model Cotera et al. (2001) used to model the HH30
near-IR scattered light images.  These grains are larger than ISM grains but
not as large as the disk midplane grains.  For the envelope we use the grain model from 
Paper I which has a size distribution that fits an extinction curve typical of the more
dense regions of the Taurus
molecular cloud, with $R_V$, the ratio of total-to-selective extinction, equal to 4.3
(Whittet et al. 2001; Cardelli, Clayton, \& Mathis 1989).   These 
grains also include a water ice mantle covering 5\% of the radius.  For the 
outflow, since small grains can condense in outflows, we use ISM grains which 
are on average slightly smaller than the envelope grains (Kim, Martin, \& Hendry 1994).  
The grain properties are shown in Figure 2, and briefly summarized
in Table 3. 
Figure 2 shows that the larger grains (e.g., the disk midplane grains shown by
the dashed line) have
have a flatter opacity, larger albedos (at longer wavelengths),
larger $g$ (average cosine scattering angle), and
lower polarization ($p_l$).

\section{Results}

This section displays spectra, colors, polarization spectra, and images for the
six evolutionary stages.  All of the sources have luminosity L=1 $L_\sun$, and are
placed at a distance of d=140 pc, for comparison to nearby low-mass star formation
regions such as the Taurus molecular cloud.

\subsection{Spectral Energy Distributions}

Figure 3 shows densities and SEDs for the 6 evolutionary states.
The smallest densities displayed ($10^{-20}$ gm cm$^{-3}$) are near the ambient density of 
$\rho = 1.67 \times 10^{-20}$ gm cm$^{-3}$ or $n_{H2} = 5 \times 10^3$ cm$^{-3}$.
In the inner regions, not shown on the scales plotted here, the densities increase to 
about $10^{-6}-10^{-5}$ gm cm$^{-3}$ in the most
dense regions of the disk
and about $10^{-11} - 10^{-13}$ gm cm$^{-3}$, depending on evolutionary state, in the inner envelope.
The SEDs in Figure 3 show the emission emergent from the entire envelope.
In nearby star formation regions, the flux is often measured in apertures that are smaller
than the envelope size.  
This affects the optical/near-IR flux the most, with smaller apertures giving less flux,
as we will show later.

For each model, ten inclinations are plotted, the top curve corresponding
to pole-on (colored pink) and the bottom to edge-on (green).  
The largest variations with inclination appear in the mid-IR, with edge-on sources
showing a broad dip near 10 $\mu$m in the Class I-II sources.  The 9.8 $\mu$m
amorphous silicate feature is much narrower than this.  The broad dip is
due to two effects:  the large extinction in the disk midplane that blocks thermal
radiation from the inner disk+envelope; and the low albedo that prevents
radiation from scattering out the polar regions. The optical extinction, A$_V$, through the
disk midplane is $> 10^7$ in the Class 0 sources and $>3\times10^5$ 
in the Class I-II sources (see Paper I, Figure 3 for a plot of A$_V$ vs. inclination
angle). 
Shortward of 10 $\mu$m, the Class I and II envelopes
allow optical/near-IR radiation to scatter out through the outflow cavities (or upper layers
of the disk in the case of the Class II source).  Thus the edge-on Class I and II sources are
double-peaked, with the short-wavelength peak due to scattering, and the long-wavelenth 
peak due to thermal emission.  

The Class II model SEDs are similar to previous models 
(Kenyon \& Hartmann 1987;
 Chiang \& Goldreich 1997; D'Alessio 
et al. 1998, 1999; Dullemond et al. 2001; Dullemond 2002;
Dullemond \& Natta 2003) except
at nearly edge-on inclinations ($i \gtrsim 70$\arcdeg),  where extinction
and scattering dominate at near- and mid-IR wavelengths.  
D'Alessio et al. (1999) include these
effects and their models show similar behavior with inclination as ours.  
Dullemond (2002) calculated the 2-D structure of disks and Herbig Ae/Be stars
and found that the inner wall puffs up and shadows outer regions in
these sources.
Since we do not compute the 2-D hydrostatic equilibrium solution, we do
not have this effect in our models.
A forthcoming paper will consider this effect in T Tauri disks (Wood et al.
2003, private communication).

The more embedded Class 0 
sources have less scattered flux at optical/near-IR wavelengths
due to higher extinction in the envelope and cavity, so the ``dip'' at 
10 $\mu$m is not
as striking in these sources.  
The Class 0 source shows little variation with inclination except for the pole-on source
and at near-IR wavelengths.
The Class III source also shows little variation with inclination because 
it is optically thin
at all wavelengths.
Examining the variation of SEDs with evolutionary state, we see a tendency for
increased shortwave (0-10 $\mu$m) flux and decreased longwave
(100-1000 $\mu$m) flux with age. 

\subsubsection{Effect of Aperture Size}

As stated previously, the SEDs shown in Figure 3 include the flux emitted by the entire envelope.
Normally, observed fluxes are integrated within a given aperture size.
Figure 4 shows SEDs of the Class 0 and I sources computed in different aperture
sizes, with 1000 AU radius apertures depicted by the solid lines and 
5000 AU apertures by the dashed lines.  
The Class II and III sources are not shown since they are smaller than either
of these apertures and thus do not vary in these apertures.
Three inclinations are shown for each model ($i=18, 56, 87$\arcdeg) in
three colors (pink, blue, green respectively).
The large aperture results show much more scattered flux at wavelengths
less than 10 $\mu$m.  
The Class 0 source has no near-IR flux in the small aperture.
This suggests that care should be taken in comparing
different sets of observations with either different aperture sizes
or different source distances. 

\subsubsection{Class I/Class II Confusion}

Figure 5 shows that a low-luminosity Class I and a high-luminosity Class II
source at different inclinations can resemble each other.
Our Class II model was scaled up by a factor of 5 to get the fluxes to
agree with the Class I model.
This gives higher far-IR flux than the Class I source but better agreement
in the near-IR where the Class I source scatters more light from its large
envelope.
The black lines correspond to a Class I source, and the grey lines are a Class II source.
The dashed lines show both sources at an inclination of $i=87$\arcdeg, or nearly
edge-on.
The solid lines show the Class I source at $i=41$\arcdeg and a Class II source
at $i=75\arcdeg$.
In both cases, the Class II source is viewed fairly close to edge-on so the
central source is obscured at visible wavelengths.  
In the left panel, the Class I fluxes are summed in the 1000 AU radius aperture,
and at right, the fluxes are summed in the 5000 AU aperture.
The smaller-aperture results do not show as much agreement.  Thus in nearby
star formation regions, it is possible to distinguish Class I and II sources
based on images and aperture photometry.  However, in far-away star formation
regions where the Class I source may not be spatially resolved, it may not
be so easy to distinguish without additional information such as polarization
(Section 3.3).

\subsubsection{Grain Properties}

Figure 6 shows the effects of varying grain properties in the different regions of the disk/envelope.
The solid lines 
show our standard models used throughout this paper, with the different grain models in 
the four different regions (Figure 1).
The grey line plots models which have the Envelope grains
throughout the disk/envelope/cavity.   
In models where the envelope emission (and scattering) dominates, we should see little difference
between the SEDs.
This is the case for the Class 0 source,
which shows a difference only at optical/near-IR wavelengths for
the pole-on source, due to the slightly higher extinction in the cavity of the
Envelope grains compared to the Outflow grains.
The late Class 0 source shows less of this effect due to lower cavity density,
but shows a difference at wavelengths longer than 100 $\mu$m for more pole-on
viewing angles.
This is because some disk emission can emerge into pole-on inclinations
at long wavelengths, and differences in the grain properties between the two
models are apparent.
The Class I through II sources show more differences at these wavelengths,
especially at pole-on inclinations.
In more evolved or pole-on
sources, more disk radiation can penetrate through the envelope.
This difference is greatest in the Class II source where emission comes only 
from
the disk, and therefore different grains in the disk have a noticeable effect 
on the longwave SED.
Edge-on, we can see differences near 10 $\mu$m in the late Class I through III
sources due to the different extinction opacities in the disk.  
The Class III models show little difference because the disk is optically thin and the multi-grain model (in black) uses the Disk Atmosphere grains
(Table 3 and Figure 2) which are not substantially different from
the Envelope grains.
Note that we have not attempted to created a realistic grain model
for a Class III source that would compute size distributions based
on shattering models.
More detailed models can be found in
Kenyon et al. (1999), 
Augereau et al. (1999), Wolf \& Hillenbrand (2003),
and Aigen \& Lunine (2003).

To turn the results of Figure 6 around, 
grain growth in the disk is not detectable in the SEDs of Class 0 sources,
due to the dominance of envelope emission.
Grain growth is detectable in Class I sources by the slope of the
longwave emission ($> 100 \mu$m) which is flatter for large grains
in a similar manner to Class II disks (Beckwith, Henning, \& Nakagawa 2000).
These results agree with those of Wolf et al. (2003) who fit near-IR images of the
IRAS 04302+2247 with ISM-like grains in the envelope, but required large grains in
the disk to fit the submm images.

\subsection{Colors and Magnitudes}


In anticipation of SIRTF observations, we show in Figures 7 and 8
colors and magnitudes for the 
near-IR J, H, and K bands; the mid-IR SIRTF IRAC bands at 3.6, 4.5, 5.8, and 8 $\mu$m; 
and the far-IR SIRTF MIPS bands at 24, 70, and 160 $\mu$m.
We also show N-band (10 $\mu$m) magnitudes for comparison to current ground-based observations.
The colored points are our models with colors corresponding to evolutionary state
as shown in the legend at top left.  The size of the circles correspond to inclination
from pole-on (large) to edge-on (small). 
In Figures 7a and 8a, the fluxes have been computed in an aperture size of
1000 AU radius,
or 7.1\arcsec at a distance of 140 pc.
Figures 7b and 8b show results for a 5000 AU radius aperture.  These plots make it
clear that in comparing results of star formation regions at different distances,
the aperture size should be taken into account.

All of the models have fluxes brighter than 1 $\mu$Jy at the wavelengths
shown except for some of the
Class 0 points, which are fainter at wavelengths smaller than 5 $\mu$m.  
For these sources, we show fluxes as faint as 0.01 $\mu$Jy to account
for the possibility of a high luminosity Class 0 source in a nearby
star forming region.  
Most of the shortwave Class 0 fluxes are too faint to show in the small aperture plots
(Fig. 7a) except at long wavelengths or pole-on inclinations.

\subsubsection{Distinguishing Protostars from Other Point Sources}

The small black symbols in Figure 7 are computed colors of a variety of different
kinds of objects based on complete 1-35 $\mu$m infrared spectra using the SKY program
(Wainscoat et al. 1992,
Cohen 1993.).
The crosses correspond to main sequence, red giant and supergiant stars;
the diamonds are AGB stars, squares are planetary nebula,
the asterisks are reflection nebulae,
the small X is an HII region and the big X is a T Tauri star (Cohen \& Witteborn 1985).
Since the SKY spectral library does not exist at wavelengths longward of
35 $\mu$m, the bottom
 right panel of Figure 7 does not show SKY results.
This Figure shows a lot of overlap between our predicted star formation colors and
those of AGB stars and planetary nebulae.  
However, the bottom left panel, [3.6]-[5.8] vs [8.0]-[24],  shows significant separation.
For the most part, the main sequence, red giant and supergiants are well
separated from the star formation models.  
The arrows show reddening vectors for the standard diffuse ISM law (in grey)
and for our Envelope dust model, more representative of dark clouds.
In the near-IR, the star formation points will overlap with reddened main sequence
and other sources (top two panels).  
However, several of the mid-IR panels
show reasonable separation from reddened galactic point sources, especially
the bottom left panel again.
The overlap with AGB stars, planetary nebulae, and reflection nebulae is
only a problem for distant star formation regions in the Galactic plane,
where we may expect to see a total of no more than 300 of these
sources per square degree.

\subsubsection{Distinguishing Protostellar Evolutionary States}

How well do the colors distinguish evolutionary states in the star formation models?
In general, the younger sources appear to be more red, though there is overlap
due to inclination effects and aperture size (Fig. 7b).
As Figure 4 showed, the Class 0 source has little measurable flux below 
$\sim 5 \mu$m for all but the pole-on source.  Thus for many of the panels, only
the pole-on inclination is shown in Figure 7a, the smaller aperture results.
In the near-IR sequences (top two panels), the pole-on Class 0 point falls in the middle of the
other models.
At longer wavelengths, the Class 0 points that are observable (in dark blue)
tend to lie in a separate region from the other sources, in the 
small aperture results (Figure 7a).
However, in the large aperture results of Figure 7b, the Class 0 sources
are in many cases {\it bluer} than the more evolved sources, even some of the Class III
sources,
especially at K-N, [5.8]-[8.0], [8.0]-[24] in Figure 7b.
This shows again the importance of considering aperture size because of the
contribution of scattered light on large scales in the envelope especially in the youngest
sources.
The Class 0 points are redder in K-[3.6], [3.6]-[4.5] but only by about 0.5-1 magnitude.
They are much redder at [24]-[70], as shown in the bottom right of Fig. 7b.

In several of the near-IR and mid-IR sequences (top four panels) notice that the
bluest sources are edge-on Class I sources.
This is true for both the small and large aperture results.
The reason for this is that the edge-on disk blocks 
all of the reddened stellar flux, leaving only the scattered flux from the 
envelope which is relatively blue
(due to decreasing scattering opacity with increasing wavelength).   

\subsubsection{Comparison to Observed Color-Color Diagrams}

The J-H vs K-[3.6] and K-[3.6] vs K-N can be compared to ground based observations
of J-H vs K-L and K-L vs K-N, for example the Taurus-Aurigae cluster at
140 pc (Wood et al. 2002b) and the NGC 2024 cluster at a distance of 415 pc
(Haisch et al. 2001).
The Taurus cloud is sufficiently nearby for our small aperture results to apply, and
indeed our results in Figure 7a agree with those shown in Wood et al. (2002b, their Figure 7).
It is interesting to note that our large-aperture K-[3.6] vs K-N plot 
includes Class 0 sources
in the same region the Haisch et al estimate Class II sources to lie.  
In the J-H vs K-[3.6] panel, the Class 0 sources also fall
in the same region that Haisch et al. estimate Class I sources to lie (their Figure 7).
And the Class I sources overlap with the Class II source region.
Thus, in more distant star formation regions, it will be more difficult to separate
evolutionary states by their colors alone.
We emphasize that scattered light, especially when measured in large apertures
or distant sources makes Class 0 and I sources appear much bluer than is perhaps
generally thought.

\subsubsection{Color-Magnitude Diagrams}

The sequences in Figure 8a, the 1000 AU radius aperture results, 
show that in general the edge-on sources are fainter, 
and the younger sources are redder. 
Note the large range in magnitude
(y-axis) for all of these sources, despite the fact that {\it all of these models have the
same luminosity}.   Even more striking
is the variation in magnitude from a given model due to inclination.  Not surprisingly,
this effect is larger for the younger sources due to their large variation of extinction
with inclination.  
As expected, the Class II and III sources
have fairly constant flux with inclination until they become edge on.
Because the edge-on sources are fainter, they could be mistaken for much lower
luminosity sources such as brown dwarfs (Walker et al. 2003).
The variation of magnitude with inclination is much smaller at 70 $\mu$m because of
the lower envelope/disk extinction and more isotropic thermal emission.  

Figure 8b shows again the overlap in colors of all the evolutionary states
when viewed through large apertures (5000 AU radius).  At J-K, the Class 0 source has the same colors
as a Class II and III source.  At [5.8]-[8], the Class 0 source is bluer than
all except the edge-on Class I and II sources.   
The [3.6]-[5.8] sequence shows more spread, but notice
that the entire evolutionary sequence ranges within 3 magnitudes.   
Since these models all have the same luminosity,
the Class 0 sources are faintest in the near to mid-IR due to large envelope extinction.
The Class III source is well-separated at [24]-[70] and 
[70]-[160].

Since the lifetimes of Class II sources are 10-100 times longer than Class 0 and
I sources, we expect observed color-color and color-magnitude diagrams to be dominated
by Class II and III sources.  Since 90\% of these sources are viewed at unobscured
inclinations, their fluxes can be used to estimate luminosity.  The caution
is that there is perhaps a 15\% contamination in these diagrams due to younger
sources (Class 0 and I) that appear blue because of scattering. 

\subsection{Polarization}

Figure 9 shows polarization plots for four of the evolutionary sequences.  
The Class 0 models did not have enough signal-to-noise to show the spectral results 
(current observing technology would probably suffer similar difficulties.)
The polarization of the Class III model is less than 0.2\% so is not shown.
In these models, the polarization arises from scattering from spherical grains in
an asymmetric envelope geometry.  
There is no component due to aligned grains (see Whitney \& Wolff 2002 for models
of scattering from aligned grains, and Aitken et al. 2003 for models that include emission from aligned
grains).
In the Class 0 and I sources,
more radiation escapes the polar regions than the equatorial, giving rise to
polarization with a position angle oriented perpendicular to the rotation axis (the $z$-axis).  
In the Class II sources, most of the scattering takes place in the disk plane.  Viewed
at most inclinations the polarization is oriented perpendicular to the disk plane, or parallel
to the rotation axis.  We define $Q=Pcos(2\chi)$, where $\chi$ is measured from the
rotation axis.  
Because $\chi$ is either 0\arcdeg or 90\arcdeg in these axisymmetric models,
we can display it as $Q$ which is either negative or positive.
For Class 0 and I sources, the $Q$ polarization is mostly negative and for Class II
sources, it is mostly positive as shown in Figure 9.
The polarization is highest towards edge-on inclinations where the unpolarized central
stellar flux becomes more extincted and the envelope scattering geometry is most
asymmetric.

The wavelength dependence of the polarization is due to several grain properties,
especially the opacity, scattering albedo and polarization efficiency.  The polarization
spectrum of the late Class 0 source is inversely correlated with the envelope extinction
(or dust opacity, see Figure 2, top left).  The more the central unpolarized flux is extincted,
the higher is the polarization of the emergent radiation.  Thus the polarization rises across 
the 3.1 $\mu$m water
ice feature and the 9.8 $\mu$m silicate feature.  Between 3 and 8 $\mu$m it decreases
due to decreasing envelope extinction as well as decreasing scattering albedo.  A competing 
effect is the polarization efficiency (Figure 2, bottom right) which increases from 1 to 10 $\mu$m.
In the late Class 0 source, the envelope is still opaque at 10 $\mu$m, so even though the
scattering albedo is very low, there is essentially no unpolarized stellar flux to dilute the
faint scattered flux.  
Thus the polarization is higher at 10 $\mu$m than at any other wavelength due to
the increased polarization efficiency.
In practice, measurement of polarization at this wavelength may be difficult due to the
correspondingly low fluxes.

In the Class I source, the polarization increases at 10 $\mu$m for the edge-on case where the
disk extincts all stellar flux, but it is not as high as at 1 $\mu$m.  This is because the
envelope scattering geometry is more asymmetric at 1 $\mu$m than at 10 $\mu$m.
At 10 $\mu$m the emerging flux comes from a wider angular range due to lower 
overall extinctions.
In the late Class I and Class II edge-on sources, the polarization is large and positive ($Q>0$)
over a large range in wavelength, peaking near 9 $\mu$m.  At these wavelengths, most
of the scattered (polarized) light emerges from the disk plane.  
The polarization is negative ($Q<0$) at visible wavelengths in the edge-on Class II source due
to a higher amount of scattering from the ambient cloud in the polar regions.
Since our disk grains
do not include the water ice features, there is no feature in the polarization spectrum in these
models.


The large polarization values in Figure 9 suggest that near-IR polarization measurements
could provide a way to disentangle
the overlap in the colors of sources of different evolutionary states, 
especially in the large-aperture results.
Figure 10 shows the same results as in Figure 7b except that the size of the
symbols correspond to the K-band polarization of that source computed in the
same aperture size as the flux.
It is clear from this figure that the Class 0 sources can be
separated from the other evolutionary states on the basis of their K-band polarization.
The edge-on Class I and II sources also show measurable (5-10\%) polarization.
Thus, we can see that if the very blue sources also have significant polarization,
they are likely edge-on Class I and II sources.  This provides a way to distinguish
these sources from brown dwarfs.

\subsection{Images}

Figure 11 shows 3-color images of our evolutionary sequence at selected wavelength
ranges.  The images have been computed through passbands corresponding
to the Hubble Space Telescope (HST) NICMOS near-IR filters F110W, F160W, and
F205W with approximate central wavelengths of 1.0, 1.6, and 2.0 $\mu$m;
and the SIRTF mid-IR IRAC and MIPS filters at approximately 3.6, 4.5, 5.8, 8, and 24
$\mu$m.
The images are displayed as ``true-color'' images with the shortest wavelength image
in blue, the middle in green, and the longest in red.  
The near-IR images are combined in the left panels of Figure 11.  
In the middle, the SIRTF IRAC bands are combined with 3.6 $\mu$m image in blue,  4.5 and 5.8 $\mu$m images averaged together in green, and the 8.0 $\mu$m image in red.   At right, the 8.0, 24, and 70 $\mu$m
images make up the three color planes.  
The displayed spatial resolution of the images in Class 0-I sources is 0.4\arcsec FWHM,
which is finer than that expected from the SIRTF IRAC camera by about a factor of 4.
The displayed resolution of the Class II and III sources is 0.1\arcsec and 0.04\arcsec
respectively.

The images are viewed at an inclination angle of 80\arcdeg showing the inner 2000 AU
for the envelope sources (Class 0-I), 500 AU of the Class II source, and
200 AU for the Class III sources.  
The intensities are scaled to a source luminosity of 1 $L_\sun$ and a distance of 140 pc.
In Figure 11a, the minimum intensities displayed are 0.06, 0.03, and 15 MJy/sr
\footnote{1 MJy/sr = 23.5 $\mu$Jy arcsec$^{-2}$} from left to right
for the Class 0-I sources.
The peak intensities displayed are as high as $10^4$ times higher than the minimum intensities, though in many of the edge-on sources, the peak intensity is only 100-1000 times higher than the minimum intensity.
The Class 0 source shows no detectable near-IR emission at the displayed 
intensities.

The near-IR images are similar to those presented by Stark et al. (2003).   The younger sources are redder.   The Class 0 source is not detected at near-IR wavelengths at the sensitivities shown here.  As pointed out by Stark et al., the disk in the late Class I source casts a shadow much larger than the disk size itself.  Hodapp et al. (2003) see a similar feature in the object ASR 41.

The SIRTF IRAC diffuse source sensitivity is expected to be approximately 0.026 MJy/sr
at 5.8 $\mu$m and lower in the smaller wavelength bands 
(SIRTF Observing Manual).
Thus, the diffuse emission from edge-on protostars in nearby star forming
regions
should be detectable with SIRTF.  The mid-IR images (middle 
panels) tend to be blue
and the far-IR images (right) are red.  This is most striking in the Class I-II sequence.
Given that the SEDs of these sources show a large dip centered at about 10 $\mu$m,
this result is understandable.  
The diffuse emission on large scales in the mid-IR images (middle panels) is due to scattering.
The Envelope grains (Figure 2, Table 3) have enough mid-IR albedo
to give large detectable scattering nebulae in the envelope sources.  
As discussed in Paper I, models of mid-IR images should provide useful information
on the albedo and therefore grain sizes of protostars.
The mid-IR images of Class II sources will also provide a useful test for the grain
properties.  As Figure 11b shows, the disk midplane thickness varies with wavelength
due to the decrease in dust opacity with increasing wavelength.  Comparison of near-IR
and mid-IR images will provide direct information of the opacity variation with wavelength.
Modeling the brightness of the scattering nebulae over this wavelength range will provide information on the dust grain sizes.
For example, McCabe, Duchene \& Ghez (2003) have detected scattered light from the HK Tau B disk
at 11.8 $\mu$m and concluded that grain sizes of 1.5-3 $\mu$m are required to produce
the necessary scattered light.

The large-scale red emission in the far-IR images (right panels) is due to thermal emission at
70 $\mu$m.  In the far-IR images, the 8 $\mu$m scattered light is very faint, and the
24 $\mu$m emission is confined to a smaller region, mainly the disk, as indicated by
the yellow color in the center and in the Class II source (Fig 11b top right).  
We note that the spatial resolution at 70 $\mu$m shown
here is much higher than will be obtained by SIRTF ($\sim 40$\arcsec).

In Figure 11b, the minimum intensities are 1, 1, and 80 MJy/sr from left to right for the Class II source; and
0.5, 0.3, and 40 MJy/sr for the Class III source.
For the Class III source, the peak intensity from the star is
about 1000 times brighter than the displayed peak in the near-IR (left)
and mid-IR (middle) panels.  
That is, the 
dynamic range between the faintest intensity shown and the peak intensity 
is about 10
million for the near- and mid-IR images.  
The corresponding dynamic range for the far-IR image (right) is
$10^5$.
The stellar
size shown in these images has a FWHM of 0.1\arcsec.  
Thus detection of the diffuse flux would require some combination of high dynamic range,
masking of the central source, excellent
stellar point-spread-function subtraction, or adaptive optics
techniques.  
Also of interest is the fact that the ambient cloud material is nearly
as bright as the disk flux
at the near-IR wavelengths (left).
This ambient material is also seen as faint emission in the Class II source at near-IR
wavelengths at the displayed intensities.

Figure 12 shows images for more a more pole-on inclination of 30\arcdeg.
The minimum intensities plotted are the same as in Figure 12.
The near-IR and mid-IR images (left and middle panels) of the Class I-III sources are
``saturated''; that is, the central pixels have fluxes that are 1-2.5 orders of
magnitude higher than the peak displayed fluxes.
For typical observed dynamic ranges of $10^4$ and
integration times suitable for measuring the stellar fluxes,
the diffuse emission of the Class I source may be not detected.
This is demonstrated also in Stark et al. (2003).
Thus pole-on Class I sources could be mistaken for Class II sources.
Several ``flat-spectrum'' sources
may be simply low-inclination Class I sources, as suggested by Calvet et al. (1994).
However, the 24 $\mu$m and 70$\mu$m images do not require as much dynamic range to
image the diffuse emission at pole-on inclinations due to the less centrally peaked
emission.  
In the Class 0 sources, the central source
is still blocked even at 30 \arcdeg inclination.
In fact, the Class 0 sources are brighter at these lower inclinations than edge-on.
This is in contrast to the Class I sources which will reveal its faint diffuse emission best when viewed
edge-on where the bright central source is blocked.

\section{Conclusions}

We have attempted to present our models of an evolutionary sequence in a way
that can be easily compared to observations.  From examination of SEDs,
colors, polarization, and images we conclude the following:

1.  Edge-on Class I and Class II sources show ``double-peaked'' 
SEDs, with a short-wavelength hump due to scattered light, and a 
long-wavelength hump due to thermal emission (Figure 3, 6a).  
The dip in the middle is caused, on the short-wavelength side,
by the lowering scattering albedo as wavelength increases from 1 to 10 
$\mu$m allowing less scattered light; and on the long-wavelength
side, by increasing extinction (opacity) as wavelength {\it decreases} from 100 
to 10 $\mu$m, allowing less thermal emssion to escape.

2.  The long-wavelength emission in Class 0 sources arises from the opaque envelope.
In Class I sources, much of the longwave emission emerges from the disk 
(Figure 6). 
Thus, the longwave Class I
emission is sensitive to grain properties in the disk, as shown by
Wolf et al. (2003).

3.  Variations due to inclination cause overlaps in SED shapes between different
evolutionary states.
Class I and II sources of varying inclination can resemble
each other (Figure 5).
This overlap in SED behavior is most apparent at mid-IR wavelengths.  
Thus mid-IR color-color plots show large overlap between different 
evolutionary states (figure 7).
`Flat-spectrum'' sources could be intermediate-inclination ($i \sim 30-50$ 
\arcdeg)  Class I sources (Fig 6c), as proposed by
by Calvet et al. (1994).

4.  The aperture size in which fluxes and colors are computed has a 
large effect in sources surrounded by large envelopes (Class 0 and I sources).
The colors can change by 1-3 magnitudes depending
on aperture size.  
This is important to take into account when comparing different
sets of observations or clouds at different distances.  
In particular, the large aperture results give very blue colors even for 
the Class 0 sources, which in some cases can
be bluer than the Class III source (Figure 7b).

5.  Several of the model mid-IR color sequences are well-separated from
reddened main sequence, red giant and supergiant stars.  However, in most
sequences, there is overlap with AGB stars, planetary nebulae, and reflection nebulae.

6.  Polarization can aid in separating evolutionary states in color-color 
diagrams.  
The Class 0 sources have dramatically higher K-band  
polarization ($\sim50$\%) than the Class I and II sources.  
The edge-on Class I and II sources, which
tend to be very blue and faint, have between 5-15\% polarization and should be
distinguishable from brown dwarfs and other faint sources.

7.  The emergent flux from Class 0-I sources vary with inclination by several 
magnitudes (Figure 8).  
The flux also varies between different evolutionary states with the same 
luminosity.
Thus, caution is required in computing luminosity functions.
However, this would not be a problem for the unobscured Class II and III sources,
which may make up $>85$\% of the sources in a cloud.  

8.  Large scale diffuse emission in nearby edge-on late Class 0 and Class I sources should be 
detectable by SIRTF in nearby star formation regions.  
The much smaller Class II sources will easily be detected
by SIRTF though at low spatial resolution.  
These images will provide useful tests of the dust
grain properties because the image brightness will be very sensitive to 
grain albedo (and thus grain size).

\acknowledgments

We thank Mike Wolff for supplying the information in Table 3.
This work was supported 
by the NASA Astrophysics Theory Program (NAG5-8587) and
the National Science Foundation (AST-9909966, AST-9819928).  
K. Wood acknowledges
support from the UK PPARC Advanced Fellowship.  
M. Cohen thanks NASA for supporting
his participation in the GLIMPSE Legacy Science Team through JPL, under Award
\#1242593 with UC-Berkeley.





\clearpage


\begin{deluxetable}{lll}
\tablenum{1}
\tablewidth{0pt}
\tablecaption{Common Model Parameters}
\tablehead{
\colhead{Parameter} &
\colhead{Description} & \colhead{Value} }
\startdata
$R_\star$       & Stellar radius                   & 2.09 $R_\sun$ \\
$T_\star$       & Stellar temperature              & 4000 K \\
$M_\star$       & Stellar mass	                   & 0.5 $M_\sun$ \\
$L$                  & Source luminosity               & 1 $L_\sun$ \\
$\alpha$        & Disk radial density exponent     & 2.25 \\
$\beta$	        & Disk scale height exponent       & 1.25 \\
$h_0$           & Disk scale height at $R_\star$   & 0.01 \\
$\alpha_{\rm disk}$ & Disk Viscosity parameter 	   & 0.01 \\
\enddata
\end{deluxetable}

\begin{deluxetable}{lllllll}
\tabletypesize{\small}
\tablenum{2}
\tablewidth{0pt}
\tablecaption{Model Variations}
\tablehead{
\colhead{Class:} &
\colhead{0} & \colhead{Late 0} & \colhead{I} &
\colhead{Late I} & \colhead{II} & \colhead{III}
}
\startdata

Envelope infall rate ($M_\sun/$yr) & $1\times10^{-4}$ & $1\times10^{-5}$ & $5\times10^{-6}$ &
$1\times10^{-6}$ & 0 & 0 \\
Envelope mass ($M_\sun$) & 3.73 & 0.37 & 0.19 & 0.037 & $1\times10^{-4}$ & 
$2\times10^{-5}$ \\
Envelope inner radius ($R_\star$) & 7.5 & 7.5 & 7 & 7 & 7 & 50 \\
Envelope outer radius (AU) & 5000 & 5000 & 5000 & 5000 & 500 & 500 \\
Disk mass ($M_\sun$) & 0.01 & 0.01 & 0.01 & 0.01 & 0.01 & $2\times10^{-8}$ \\
Disk inner radius ($R_\star$) & 7.5 & 7.5 & 7 & 7 & 7 & 50 \\
Disk outer radius (AU) & 10 & 50 & 200 & 300 & 300 & 300 \\
Disk accretion rate ($M_\sun/$yr) & $1.4\times10^{-7}$ & $2.8\times10^{-8}$  &
$6.8\times10^{-9}$  & $4.6\times10^{-9}$ & $4.6\times10^{-9}$ & 0 \\
Disk acc. lum. ($L_\star$) & 0.036 & 0.0069 & 0.0018 & 0.0012 & 0.0012 & 0 \\
Cavity density ($n_{H2}$ cm$^{-3}$) & 10$^5$ & $6.7\times 10^4$  & $5\times 10^4$ &
10$^4$  &  $5\times 10^3$ & $1\times 10^3$  \\
Cavity opening angle (\arcdeg) & 5 & 10 & 20 & 30 & 90 & 90 \\

\enddata
\end{deluxetable}

\begin{deluxetable}{llll}
\tablenum{3}
\tablewidth{0pt}
\tablecaption{Dust Models}
\tablehead{
\colhead{Description} &
\colhead{Region} & \colhead{$r_{eff}$} & $r_V$ }
\startdata
Disk Midplane   & Inner disk ($n_{H_2} > 10^8$ cm${^-3}$)  &    0.69  & 4.9  \\
Disk Atmosphere & upper layers of disk & 0.042 & 4.1 \\
Envelope & Infalling envelope &  0.048 &  4.3  \\
Outflow & bipolar cavity &  0.026  &  3.6 \\
\enddata
\end{deluxetable}


\clearpage



\begin{figure}
\figurenum{1}
\epsscale{0.6}
\plotone{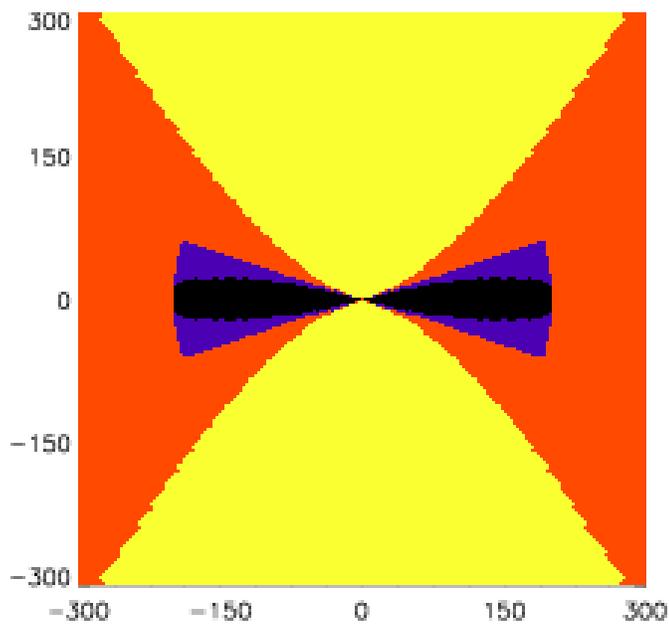}
\caption{
The inner region of the grid (300 AU) showing the regions where the different grain
models are used.  Grid cells colored black have the Disk Midplane grains (from Wood et al. 
2002); blue cells have the Disk Atmosphere grains (from Cotera et al. 2001); 
red cells have the Envelope grains (Paper I); and yellow cells have the 
Ouflow grains
(ISM grains from Kim, Martin, \& Hendry 1994).  
}
\end{figure}
\clearpage

\begin{figure}
\figurenum{2}
\epsscale{1.0}
\plotone{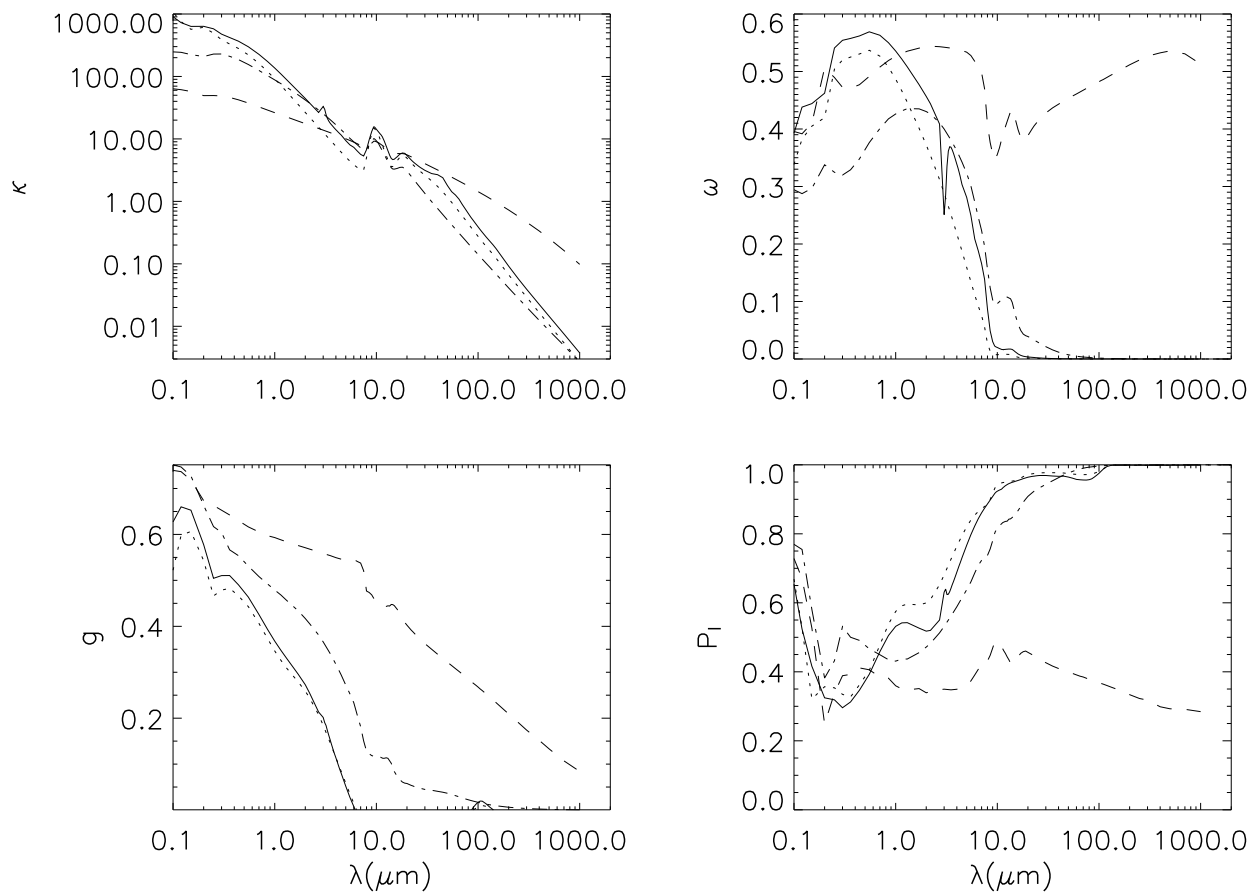}
\caption{
Dust properties of the 4 grain models.  Opacity ($\kappa$) is shown in the top left panel,
albedo ($\omega$) at top right, average cosine of scattering angle ($g$) is bottom
left, maximum polarization ($P_l$, typically at 90$\arcdeg$ scattering angle) is bottom
right.  The four lines show the four grain models from Figure 1 as follows: Envelope, solid line;
Outflow, dotted; Disk Midplane, dashed; Disk Atmosphere, dot-dashed.
}
\end{figure}
\clearpage

\begin{figure}
\figurenum{3a}
\epsscale{1.0}
\plotone{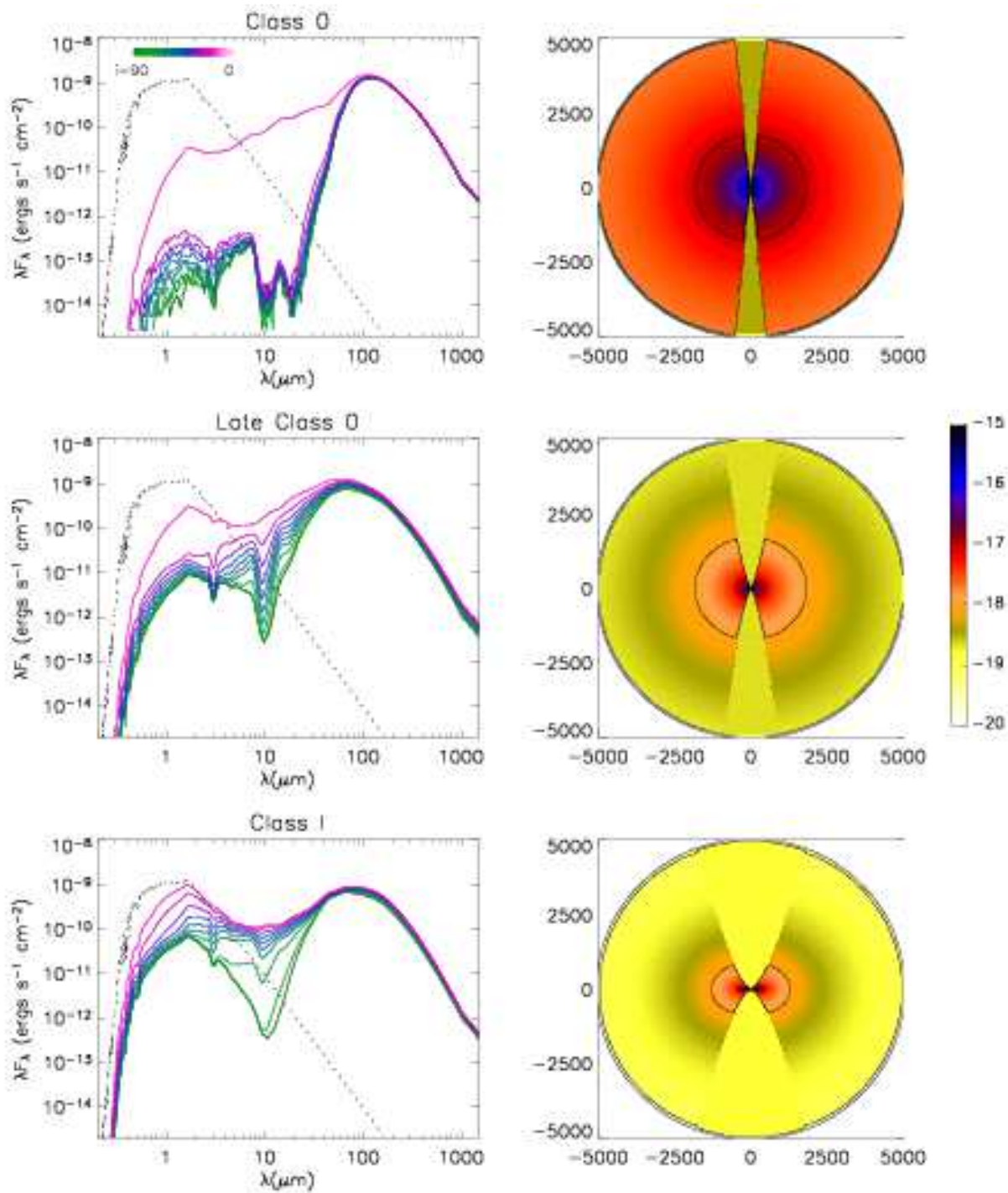}
\caption{
SEDS (left) and densities (right) the 6 models.  For the SED models, the
colors indicate inclination, as shown in the top left panel:  dark green
is edge-on, and pink is pole-on.
The density images are plotted to log scale, with the contours 
matching the 
tick marks in the color bar.
The size range is noted in the axes in AU.  Note that the Class II and
III sources (disks) are much smaller than the envelope sources.
(a) Class 0, late Class 0, \& I
(b) late Class I, Class II, \& III.  
}
\end{figure}
\clearpage

\begin{figure}
\figurenum{3b}
\epsscale{1.0}
\plotone{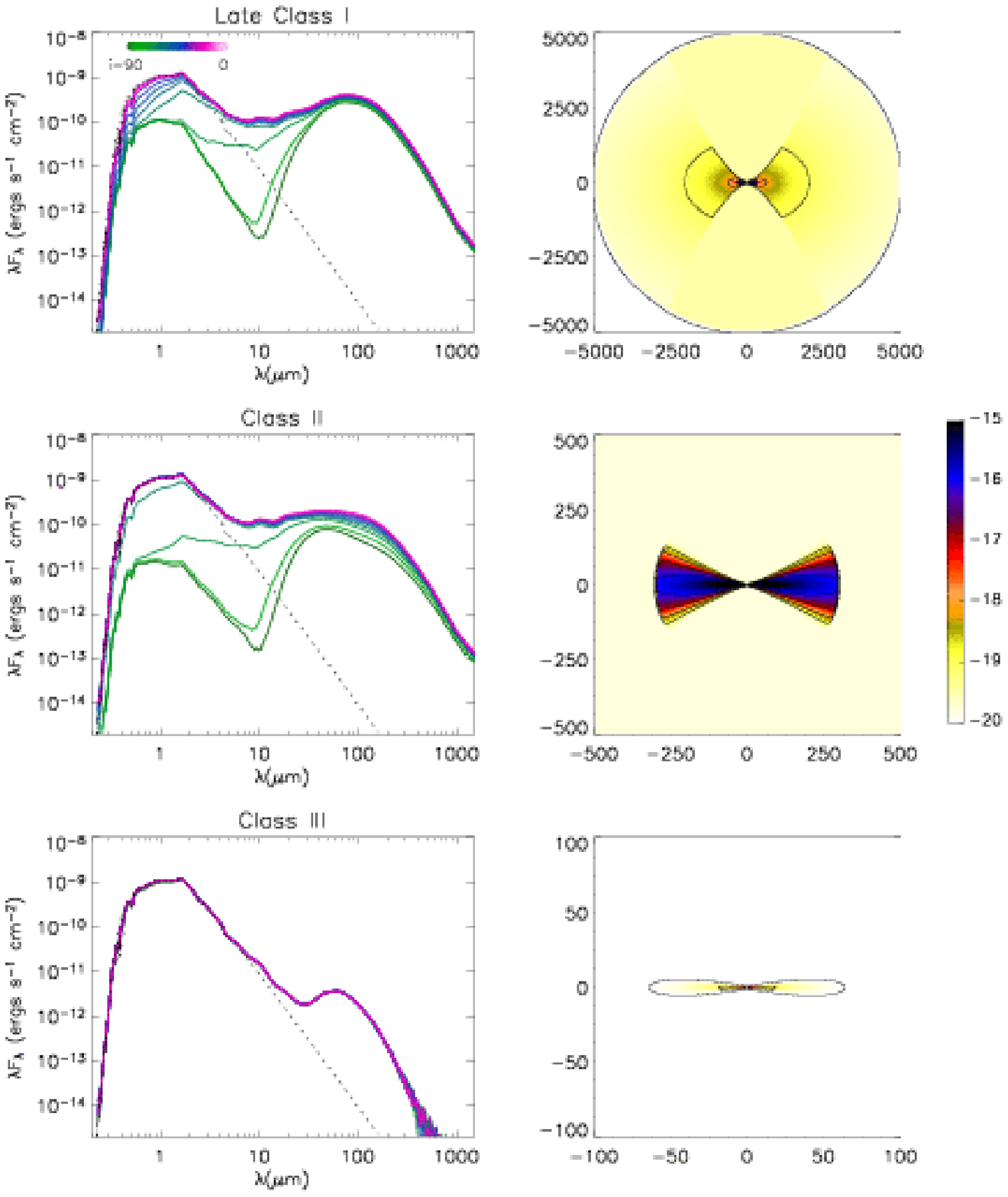}
\caption{}
\end{figure}
\clearpage

\begin{figure}
\figurenum{4}
\epsscale{1.0}
\plotone{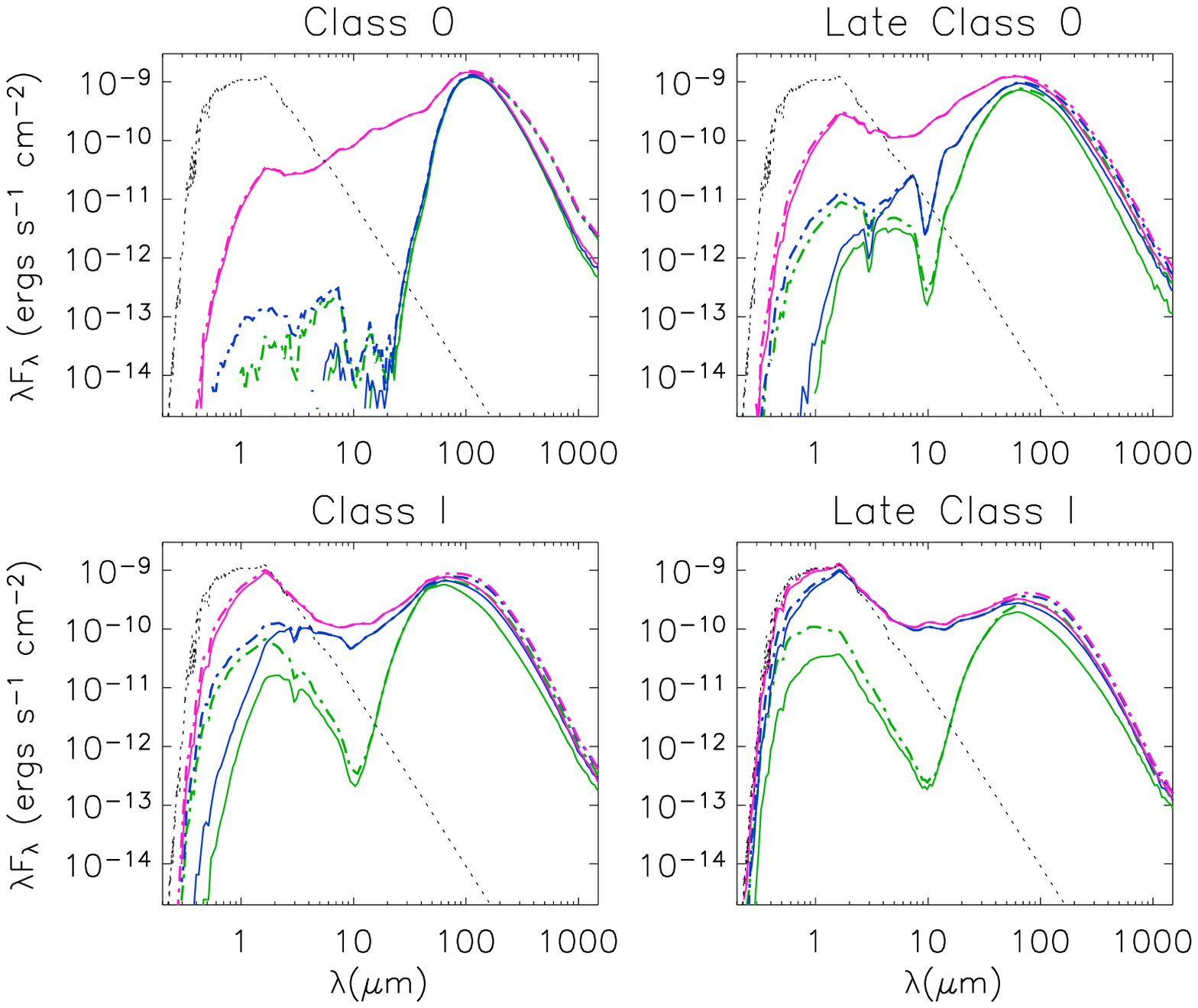}
\caption{Comparison of SEDs computed in different aperture sizes
for the sources with envelopes (Class 0-I).
Small aperture results (1000 AU radius) are shown as solid lines;
large aperture results (5000 AU radius) are dot-dashed.
Three inclinations are plotted for each model ($i=18, 56, 87$\arcdeg) as separate
colors (pink, blue, green respectively)
}
\end{figure}
\clearpage


\begin{figure}
\figurenum{5}
\epsscale{1.0}
\plotone{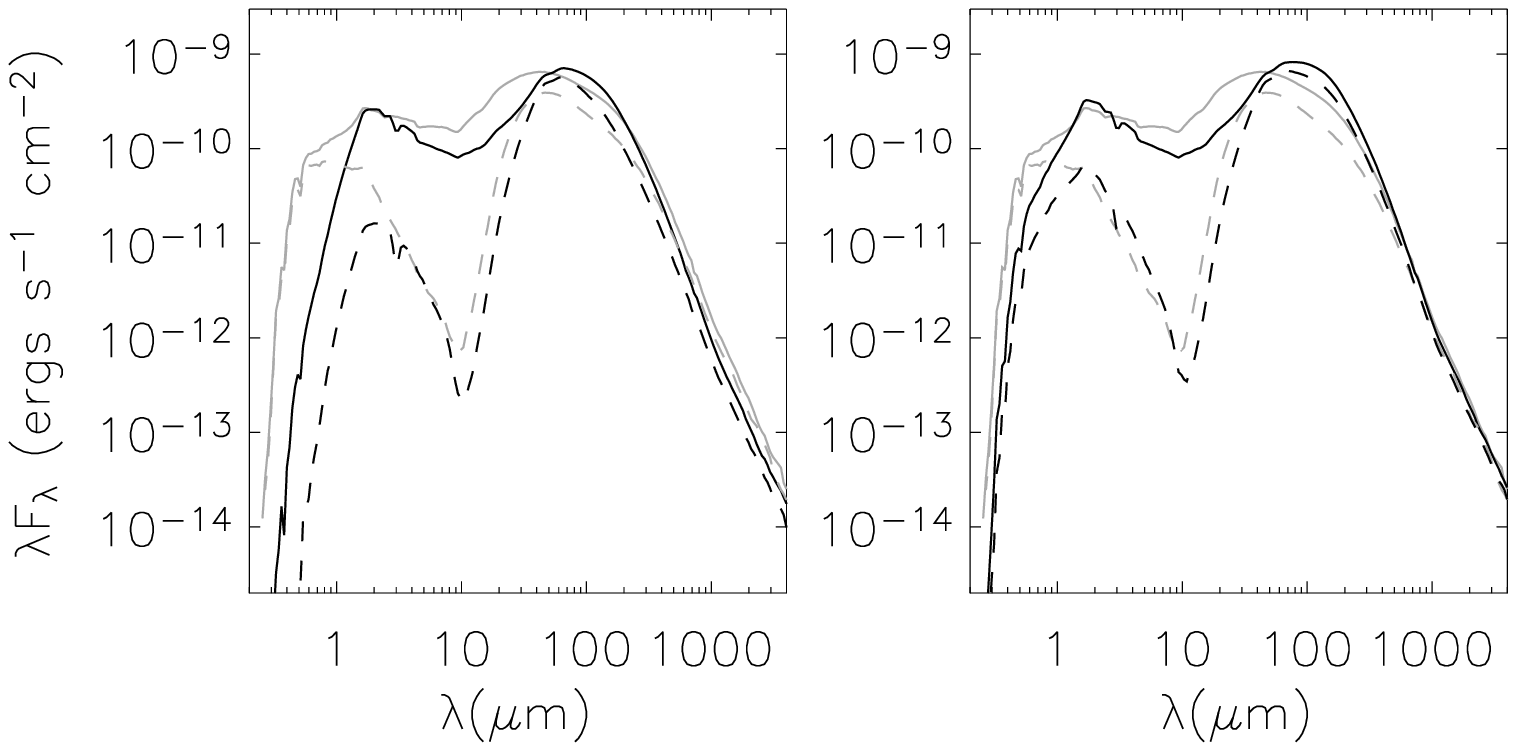}
\caption{
SEDs at selected inclinations for a Class I source (black) and Class II source (grey).
The luminosity of the Class II source is scaled up by a factor of 5.
The Class I results are shown for an aperture of 1000 AU at left, and 5000 AU at right.
The dashed lines show inclinations of $i=87\arcdeg$ for for the Class I and II sources.
The solid lines show inclinations of $i=41\arcdeg$ for the Class I source and
$i=75\arcdeg$ for the Class II source.
}
\end{figure}
\clearpage

\begin{figure}
\figurenum{6a}
\epsscale{1.0}
\plotone{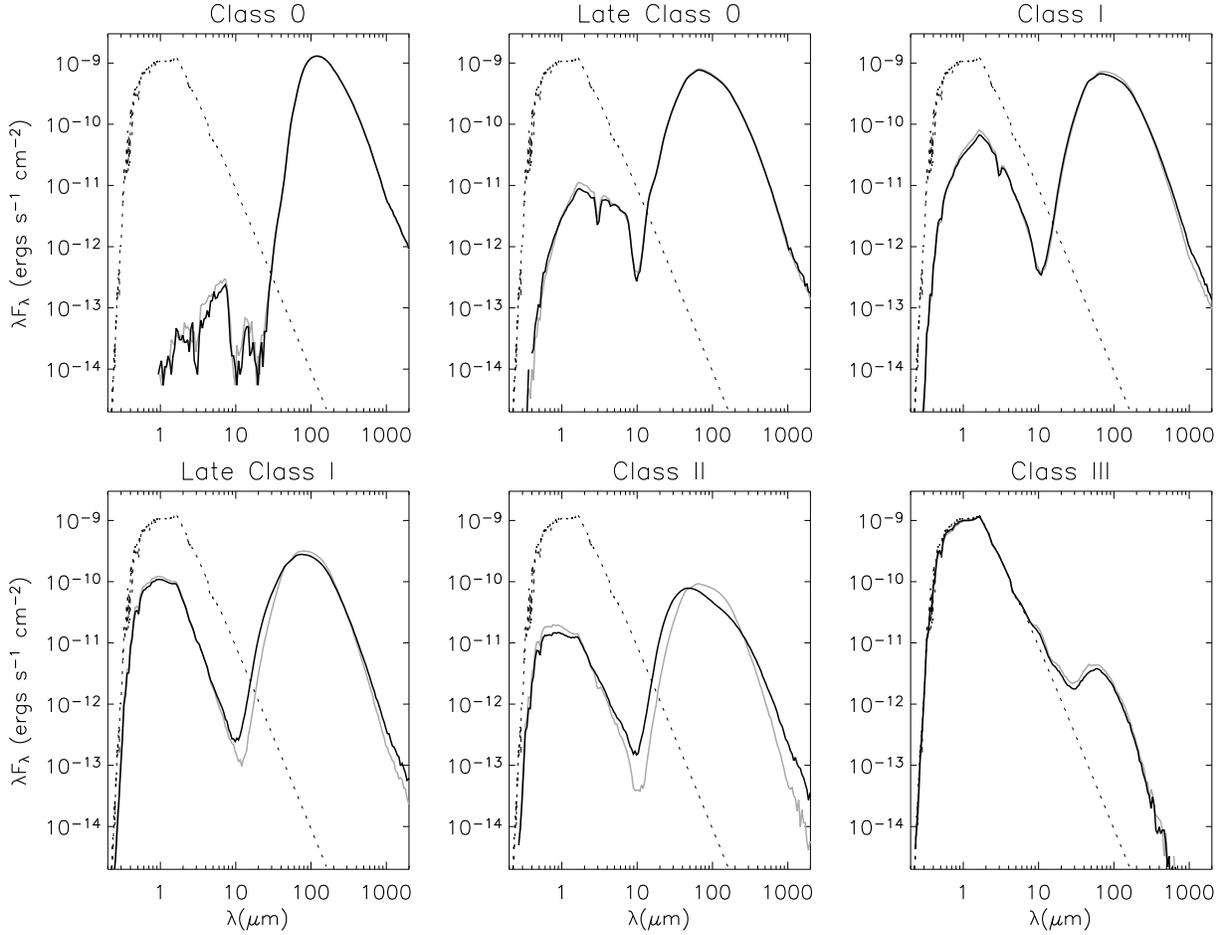}
\caption{
Comparison of the SEDs produced from the different dust models.
The black line plots the results from using different dust models for the four different regions (see Figure 1 and Table 3).  The grey line shows results using the Envelope dust ($r_v=4.3$)
throughout.
(a) inclination is i=87 (edge-on) 
(b) inclination is i=56
(c) inclination is i=18 (pole-on) 
}
\end{figure}
\clearpage

\begin{figure}
\figurenum{6b}
\epsscale{1.0}
\plotone{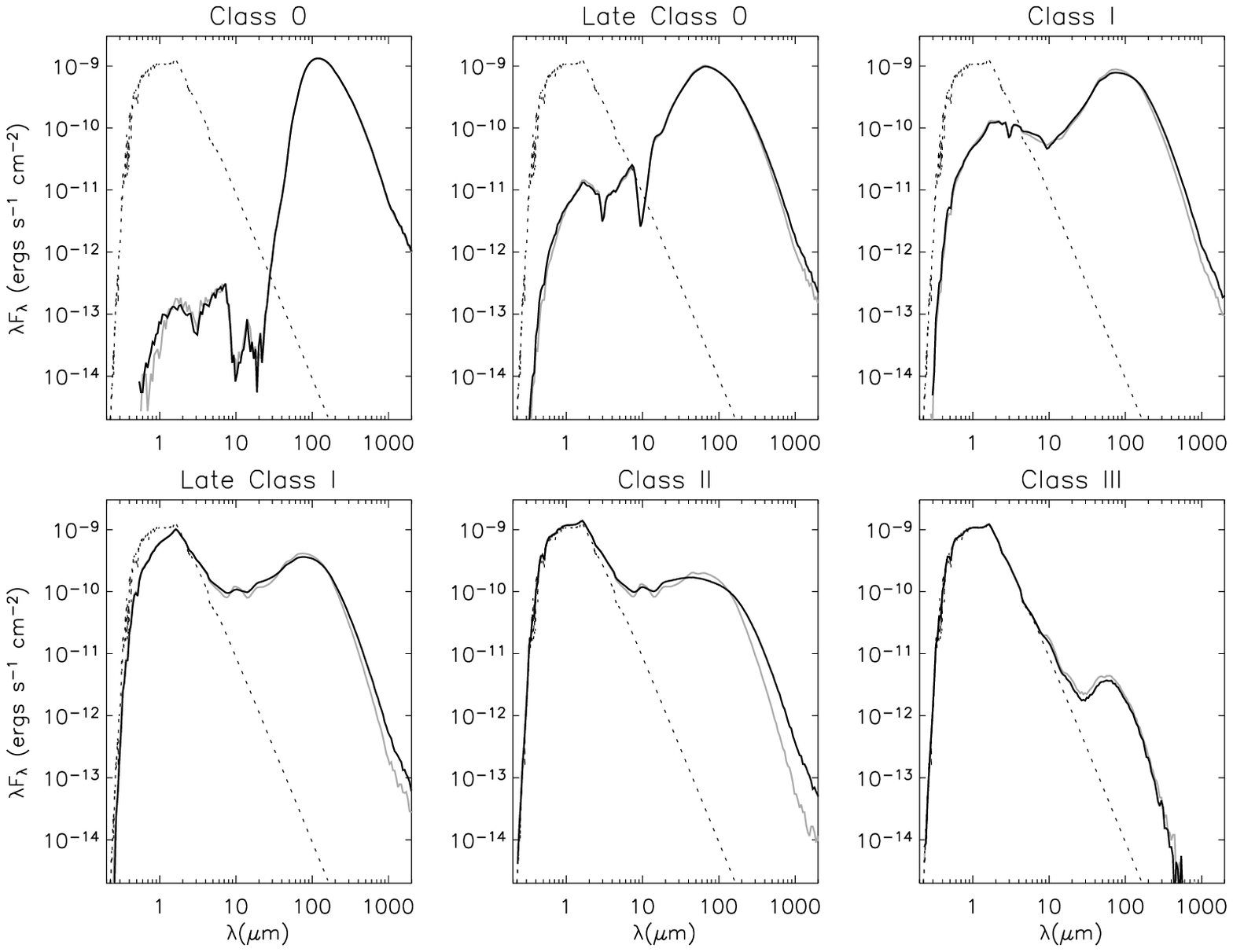}
\caption{}
\end{figure}
\clearpage

\begin{figure}
\figurenum{6c}
\epsscale{1.0}
\plotone{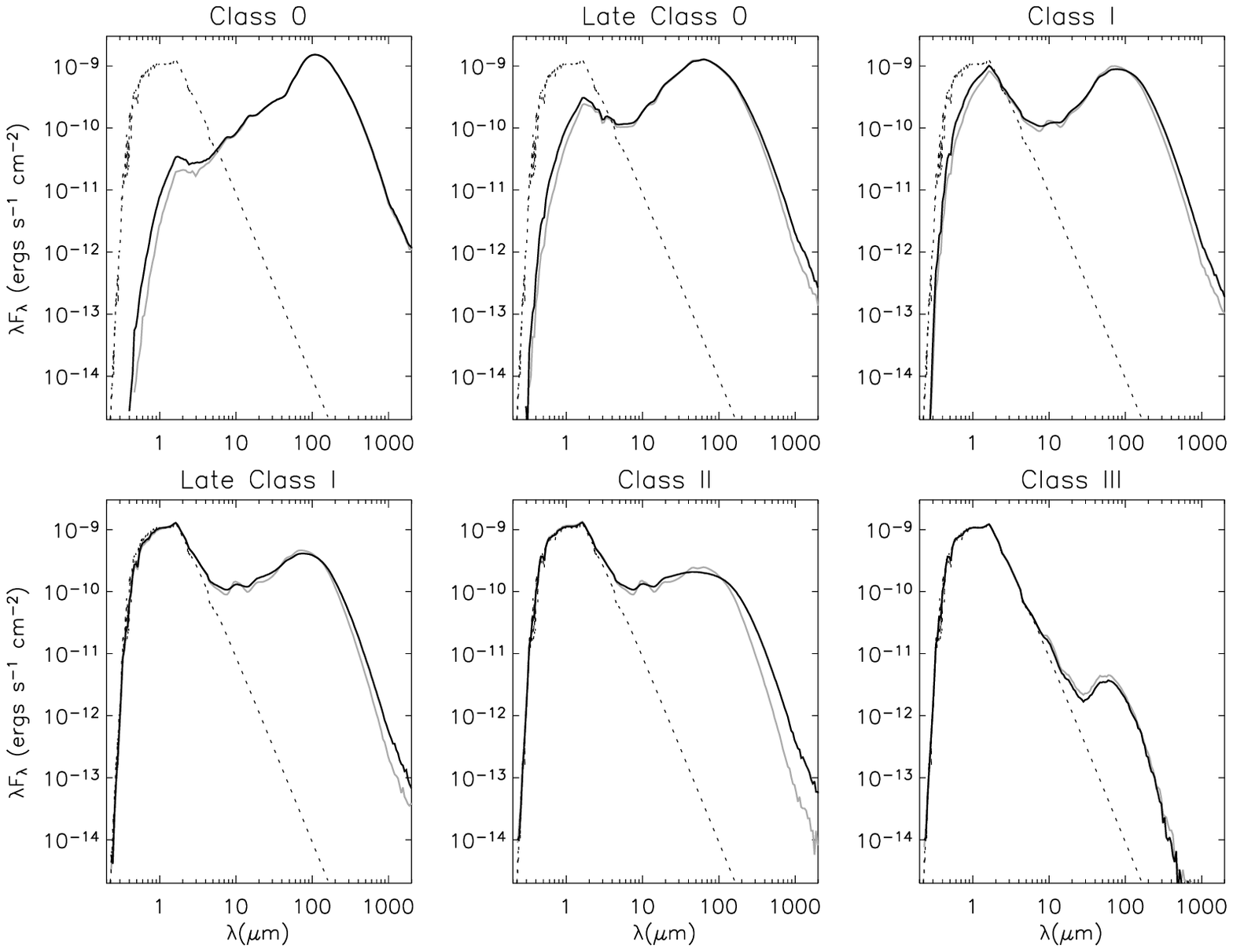}
\caption{}
\end{figure}
\clearpage

\begin{figure}
\figurenum{7a}
\epsscale{0.85}
\plotone{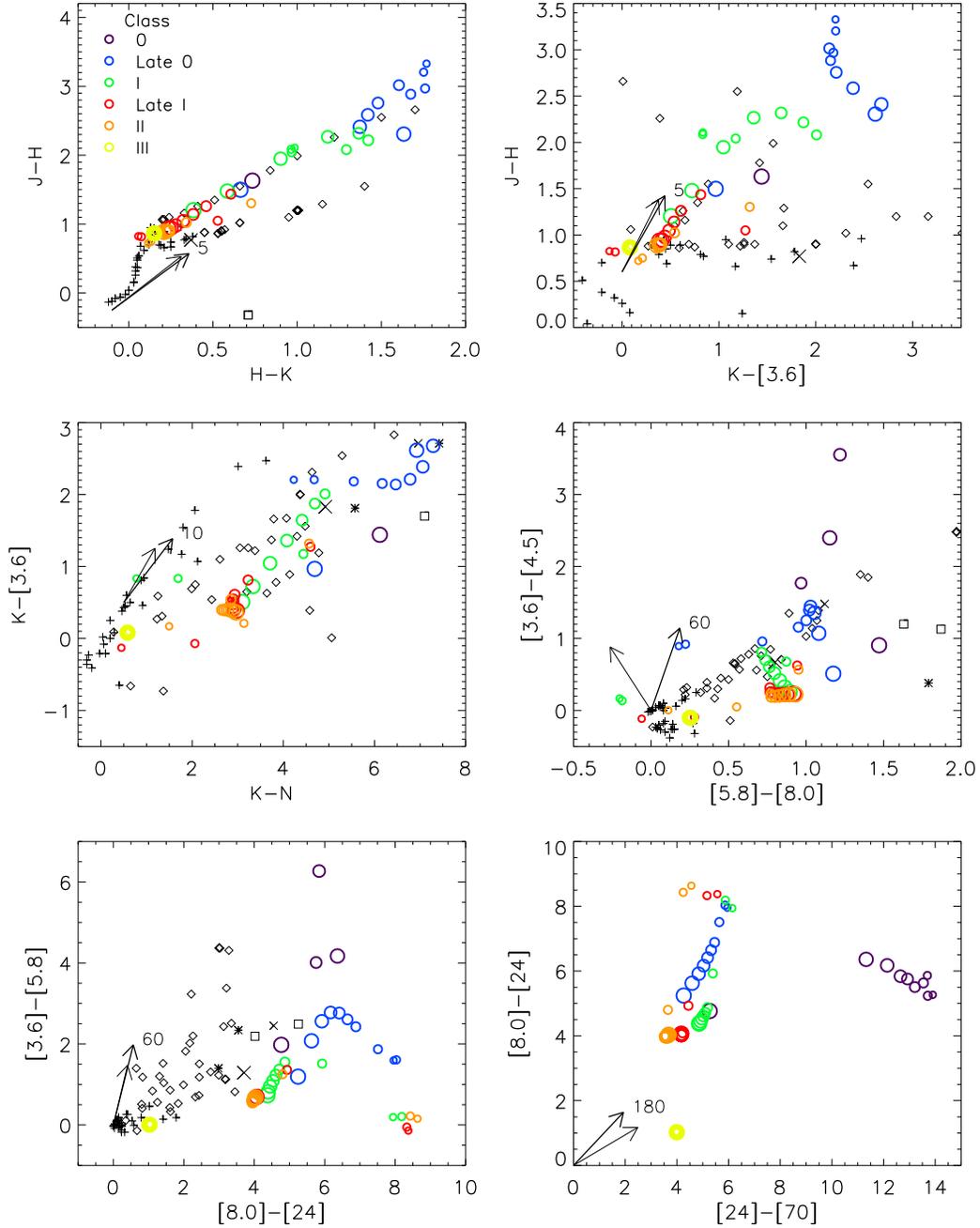}
\caption{
Color-color plots.  The legend in the top left panel shows the color correspondence
with the evolutionary sequence.  The symbol sizes vary with
inclination with edge-on the smallest.  
Arrows show the reddening vectors for standard ISM dust (grey arrow)
and dust more representative of dark clouds with $r_v = 4$ (our Envelope
grains) (black arrow).
The black symbols are predictions for other
stellar types as follows--crosses:  main sequence, red giant, supergiants; 
diamonds: AGB stars; squares: planetary nebulae; asterisks: reflection nebulae;
small x: HII region; large X: T Tauri star.
a) Aperture size is 1000 AU radius.  
b) Aperture size is 5000 AU radius.
}
\end{figure}

\clearpage
\begin{figure}
\figurenum{7b}
\epsscale{0.85}
\plotone{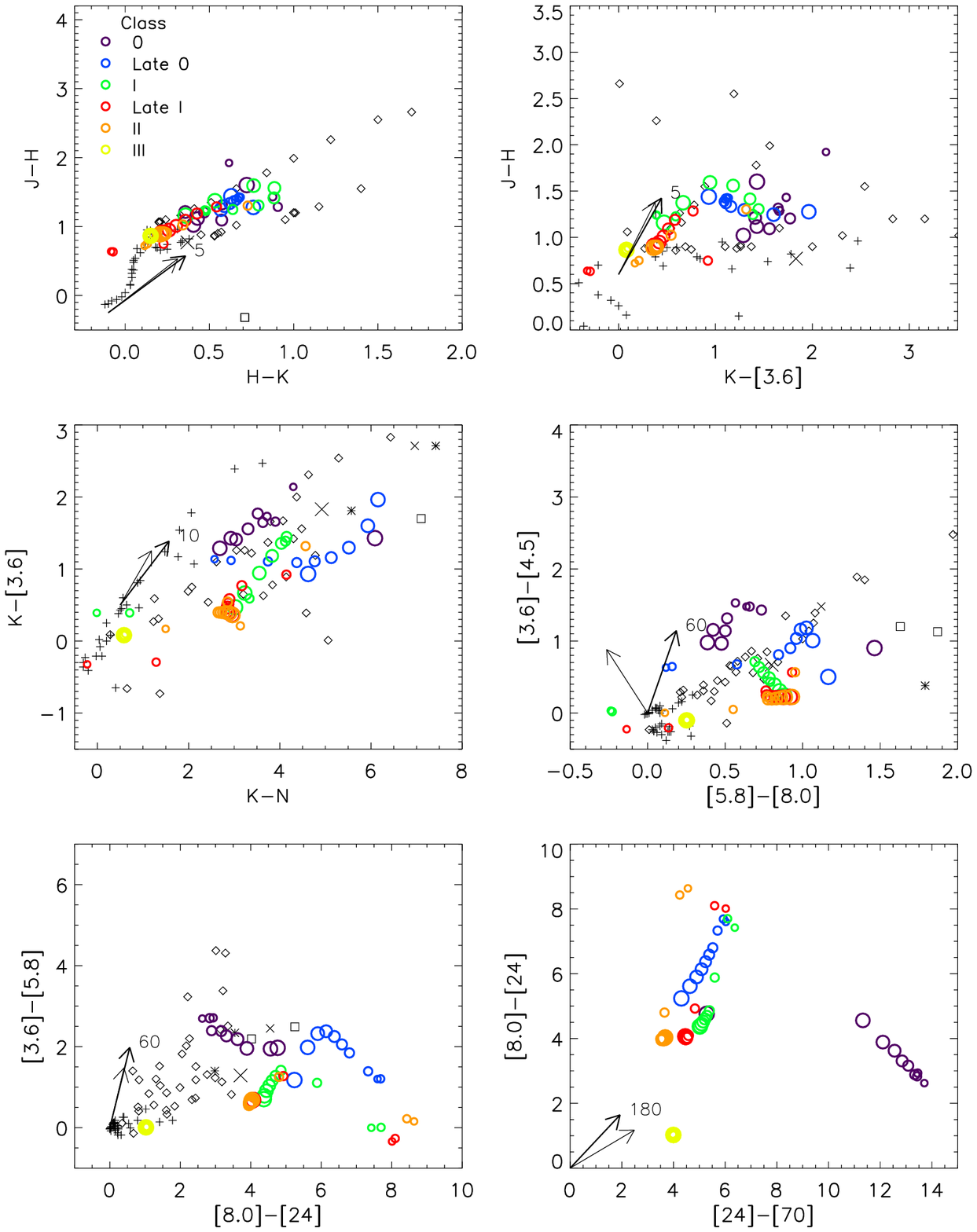}
\caption{
}
\end{figure}

\begin{figure}
\figurenum{8a}
\epsscale{0.85}
\plotone{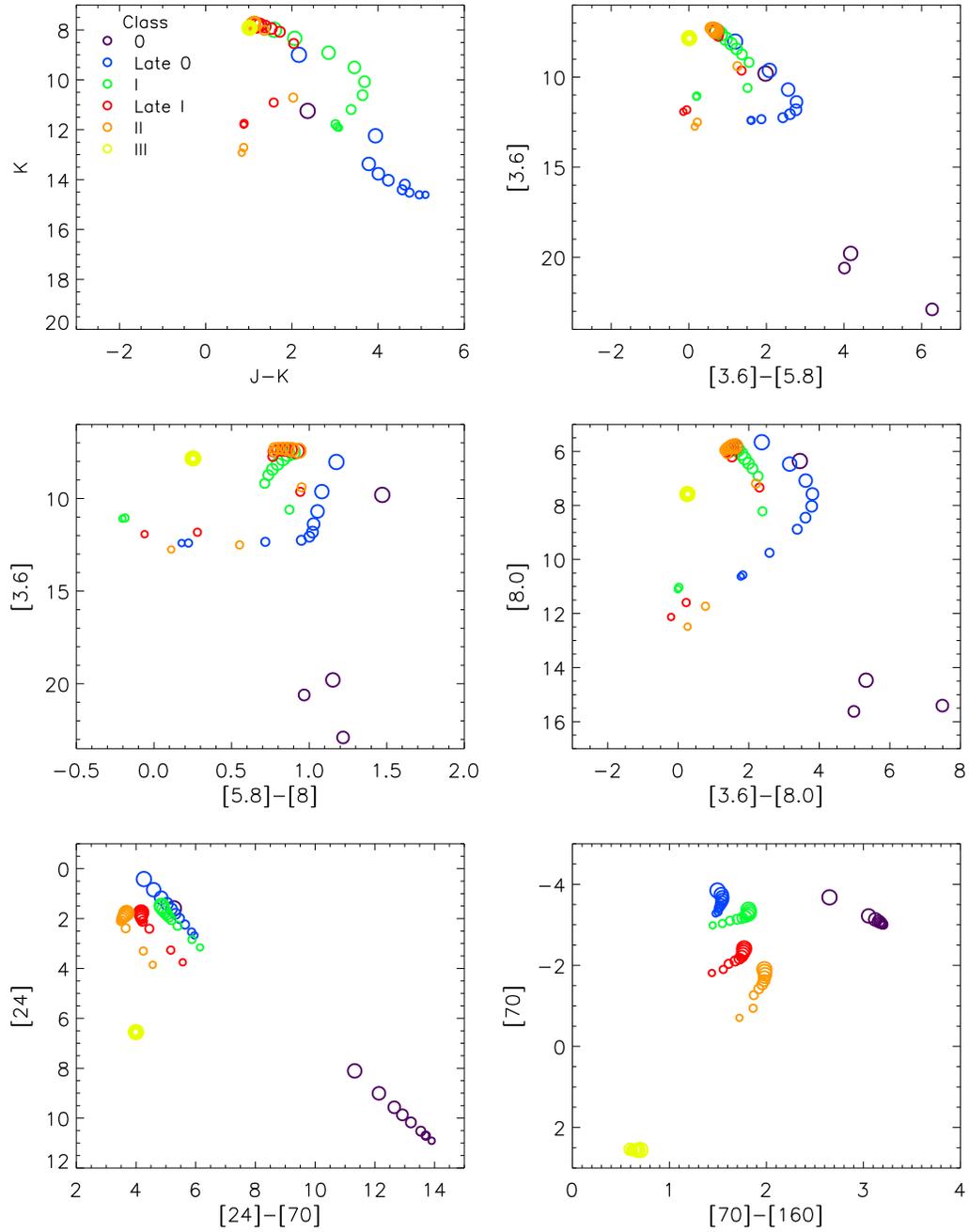}
\caption{
Color-magnitude plots.  The legend in the top left panel shows the color correspondence
with the evolutionary sequence.  The symbol sizes vary with
inclination with edge-on the smallest.  
a) Aperture size is 1000 AU radius.  
b) Aperture size is 5000 AU radius.
}
\end{figure}
\clearpage

\begin{figure}
\figurenum{8b}
\epsscale{0.85}
\plotone{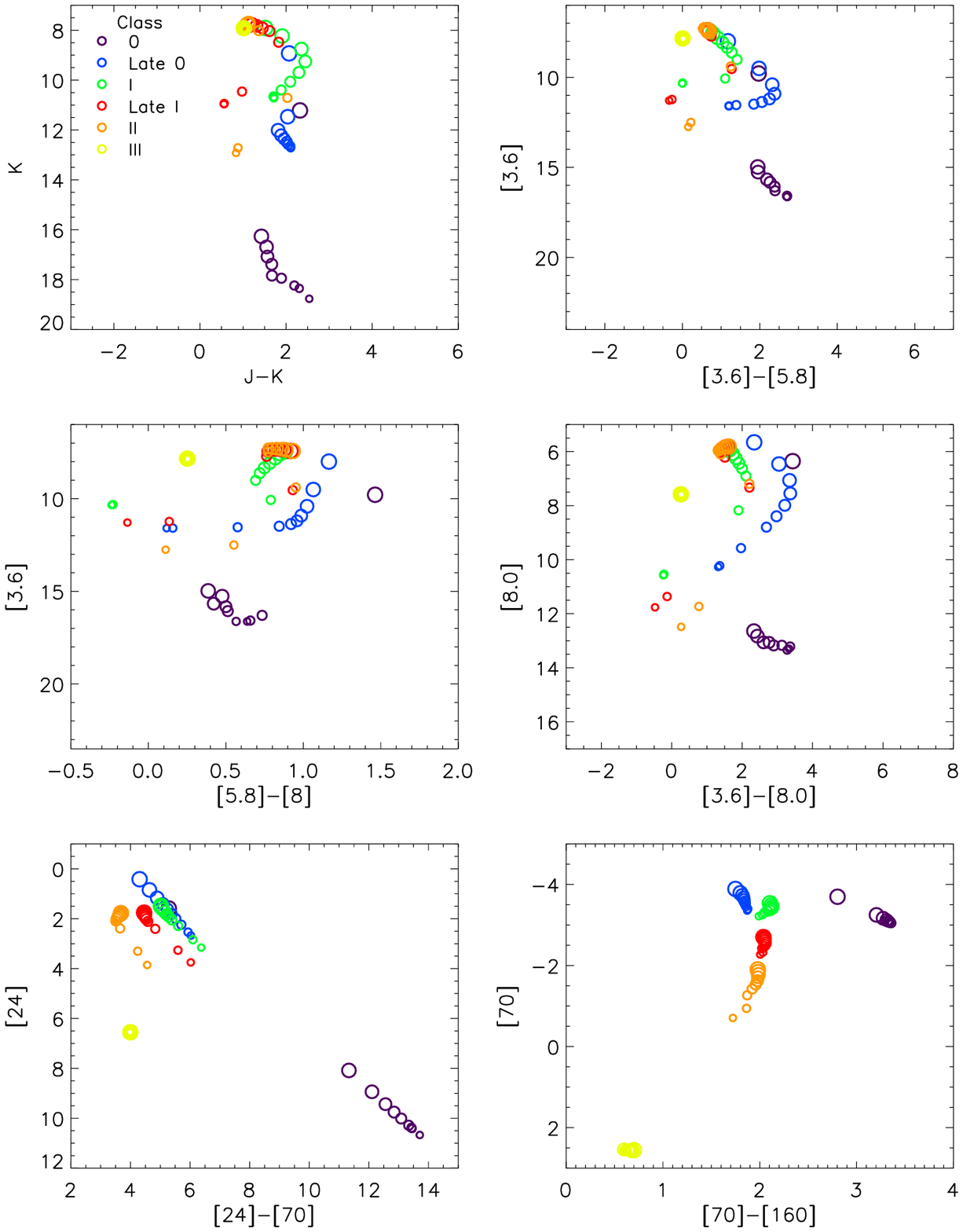}
\caption{
}
\end{figure}
\clearpage

\begin{figure}
\figurenum{9}
\epsscale{1.0}
\plotone{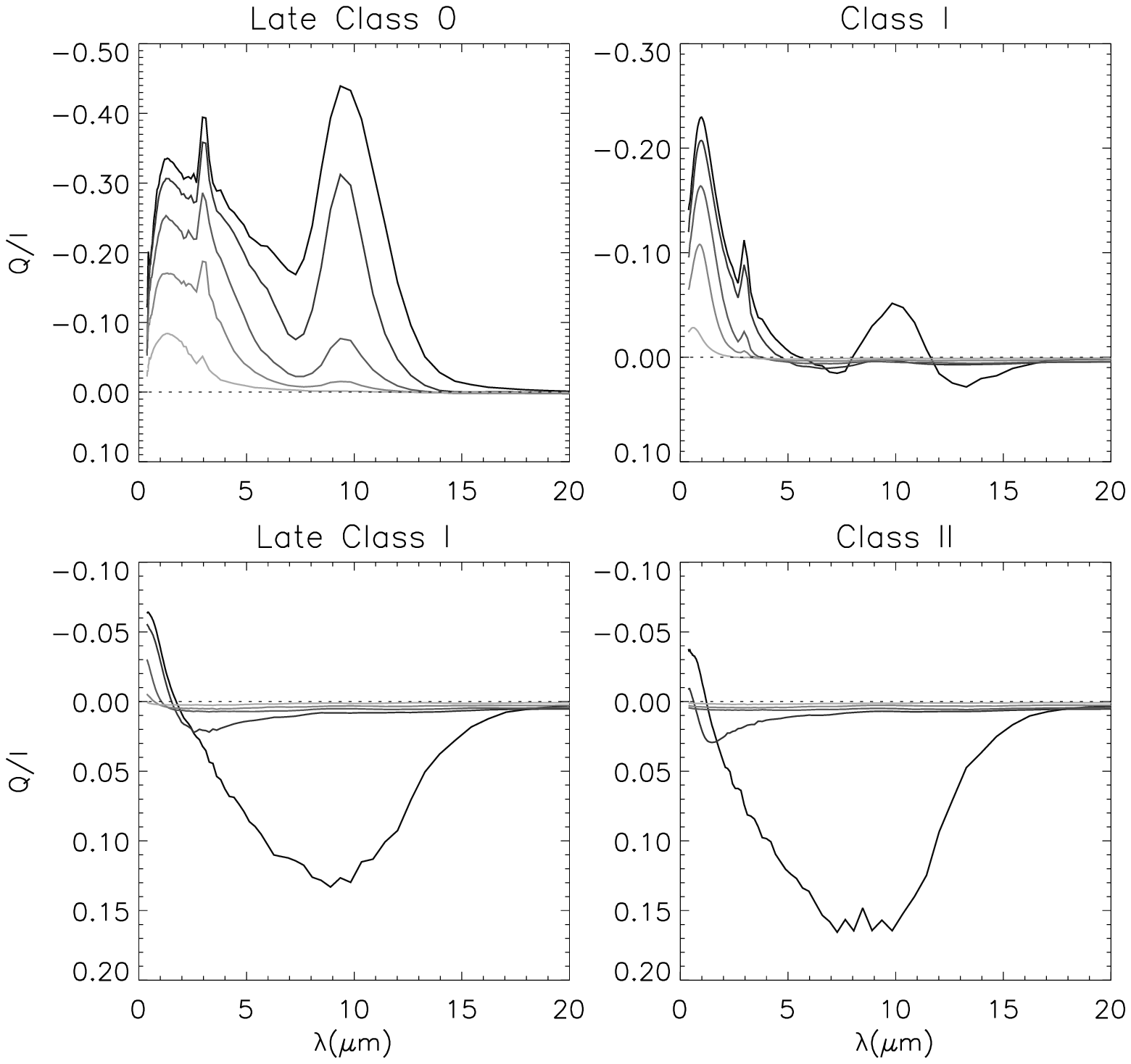}
\caption{Polarization.  Five inclinations are
plotted from $\cos i=0.05$ to 0.85 in intervals of 0.2.  The lines are
lighter grey at lower inclinations (more pole-on).
}
\end{figure}
\clearpage


\begin{figure}
\figurenum{10}
\epsscale{0.85}
\plotone{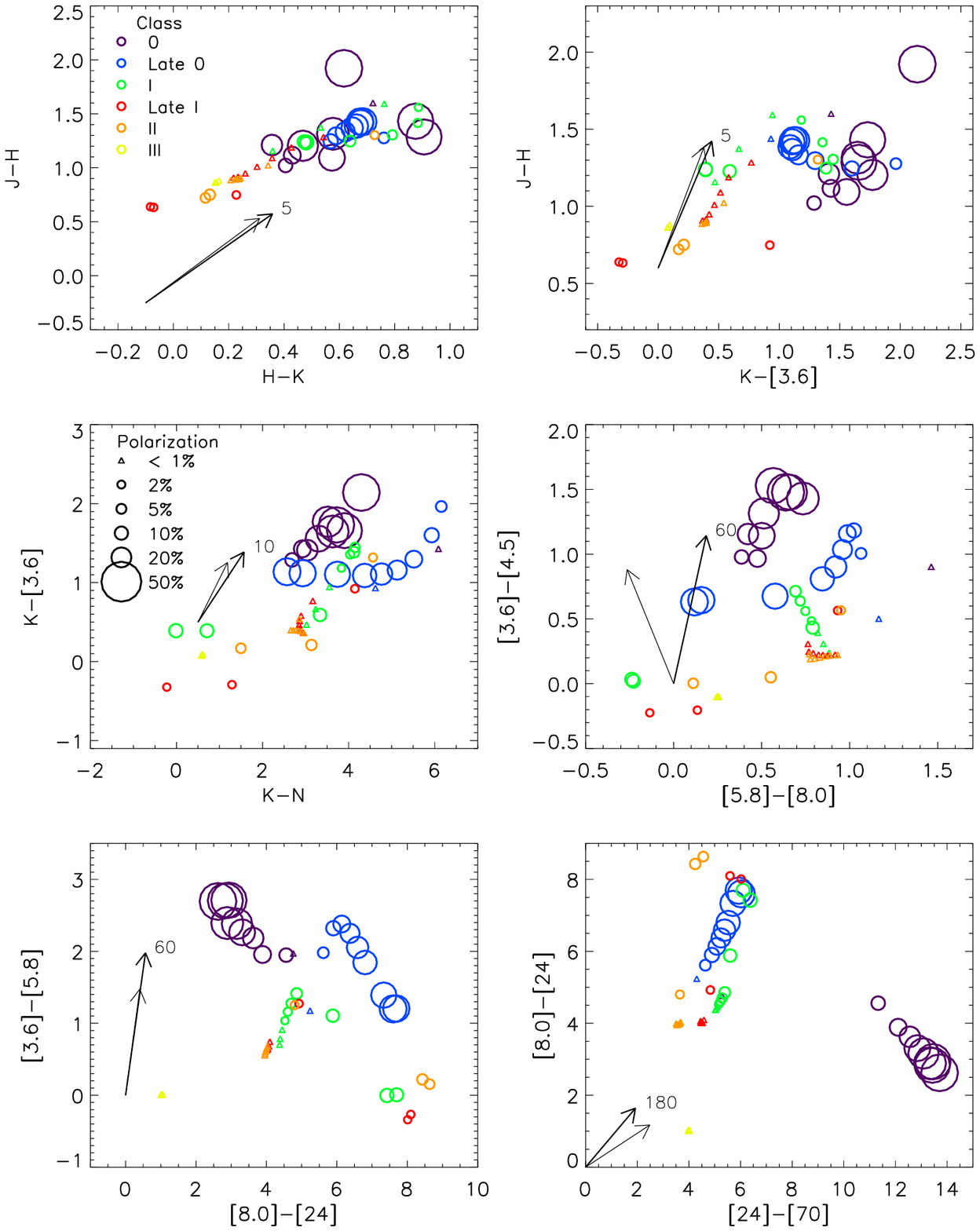}
\caption{
Color-color plots showing the large aperture results (5000 AU).  This is similar 
to Figure 7b except that the size of the symbols corresponds to the K-band polarization
as shown in the legend in the middle left panel.  Points with polarization less
than 1\% are plotted as small triangles.  This shows that near-IR polarization
can separate the Class 0 sources from others in the same
region of color-color space.
}
\end{figure}

\begin{figure}
\figurenum{11a}
\epsscale{0.95}
\plotone{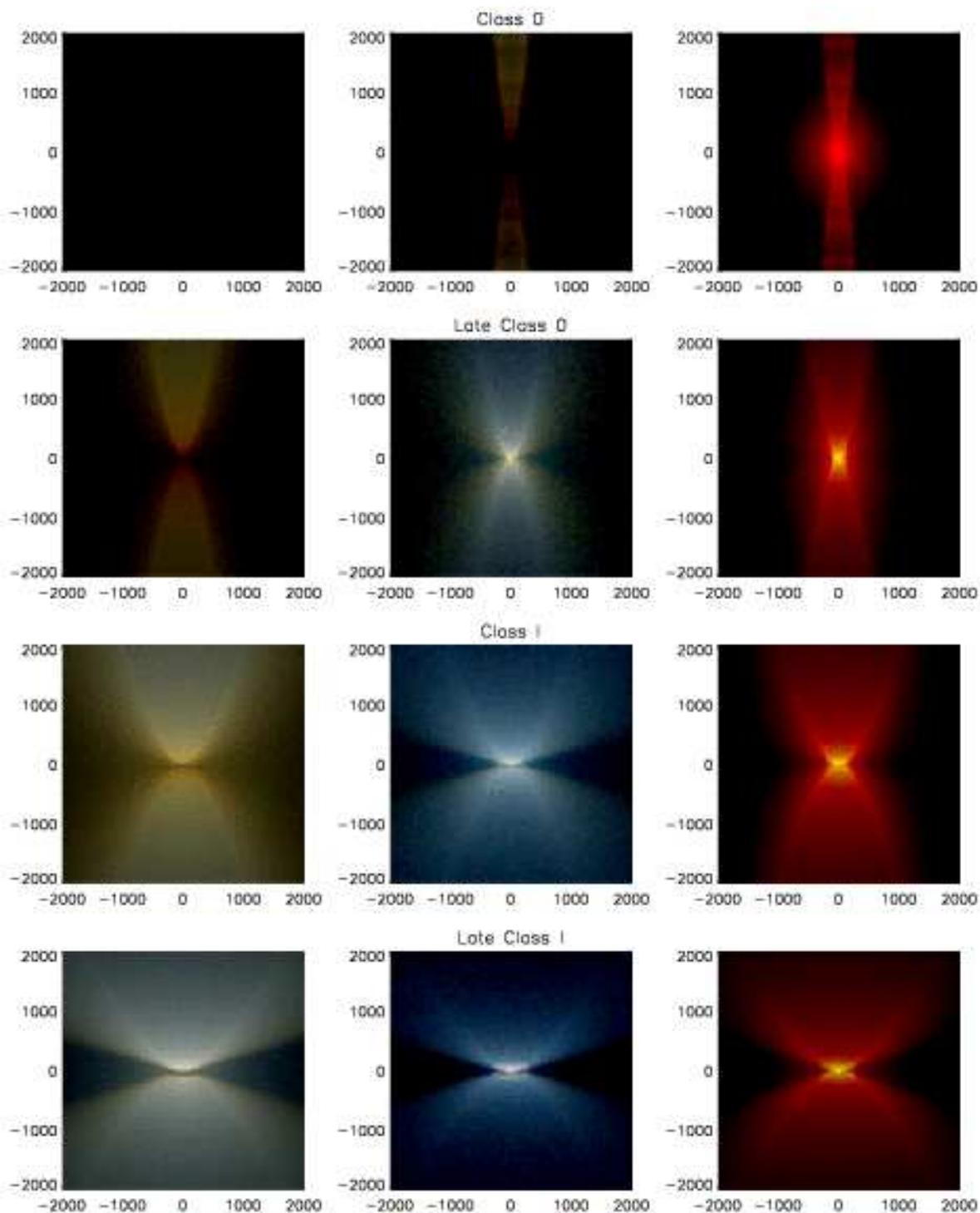}
\caption{3-color composite
images of the models.  Left: 1.1, 1.6, and 2.05 $\mu$m; middle: 3.6, average of 4.5 and 5.8, and 8.0 $\mu$m; right: 8.0, 24, and 70 $\mu$m.  
The images are scaled logarithmically from the minimum intensity described in the text
to 4 orders of magnitude above. 
The inclination is 80$\arcdeg$.
The axis labels show the size of the image in AU.  
(a) Class 0, Late 0, I, and Late I models.
(b) Class II and III models.
}
\end{figure}
\clearpage

\begin{figure}
\figurenum{11b}
\epsscale{0.95}
\plotone{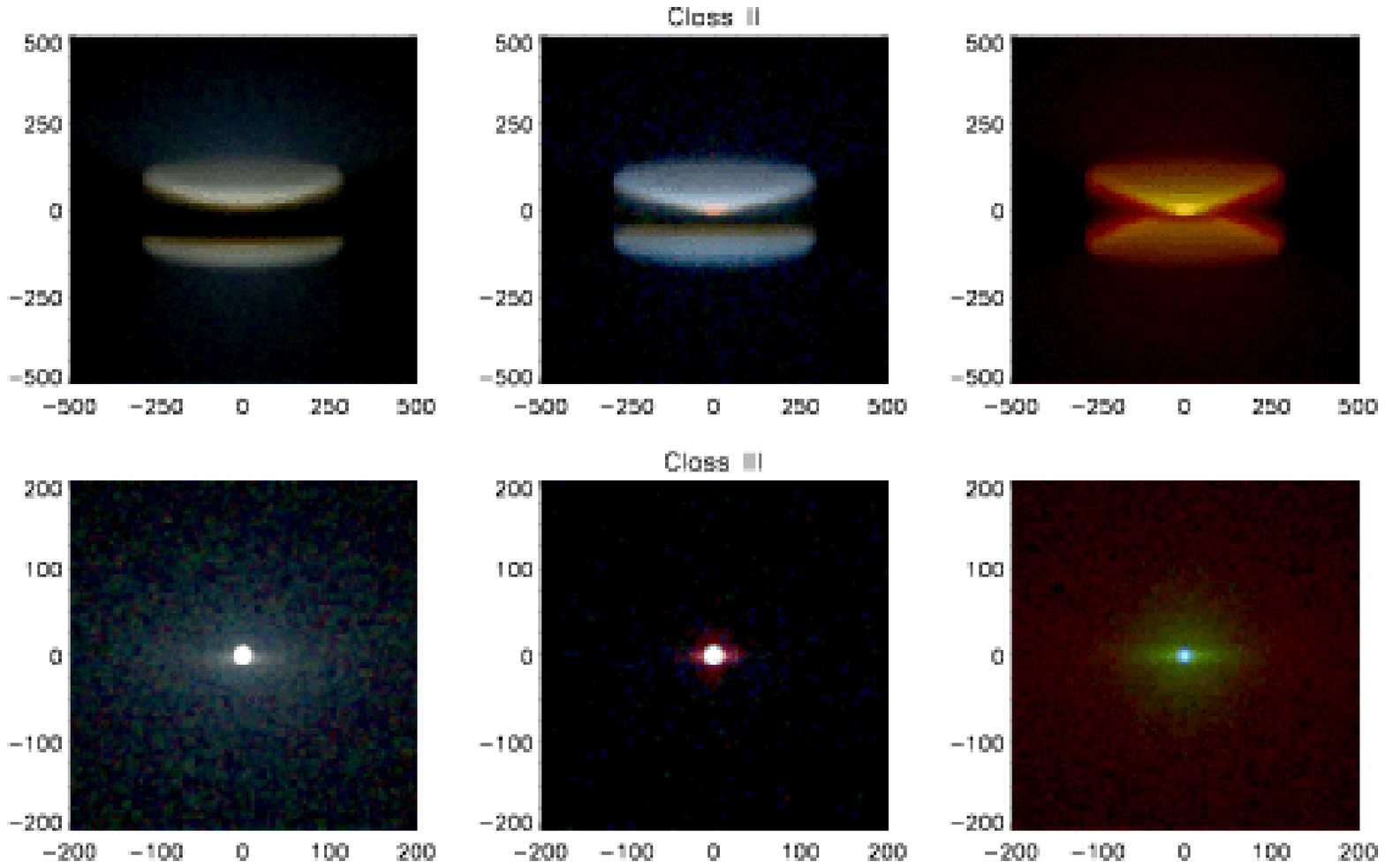}
\caption{}
\end{figure}

\begin{figure}
\figurenum{12a}
\epsscale{0.95}
\plotone{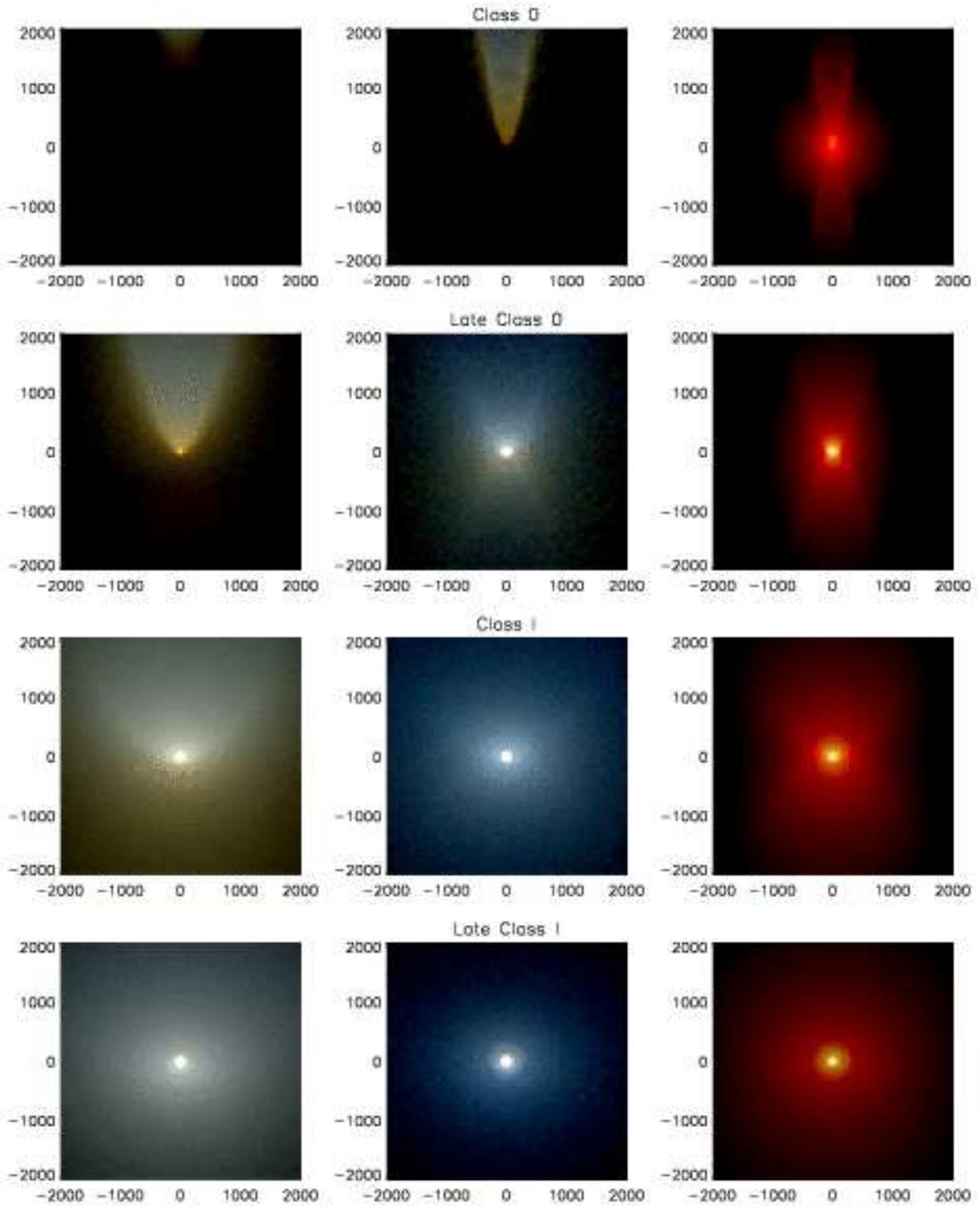}
\caption{Same as Figure 12 but viewed at a 30\arcdeg inclination
}
\end{figure}
\clearpage

\begin{figure}
\figurenum{12b}
\epsscale{0.95}
\plotone{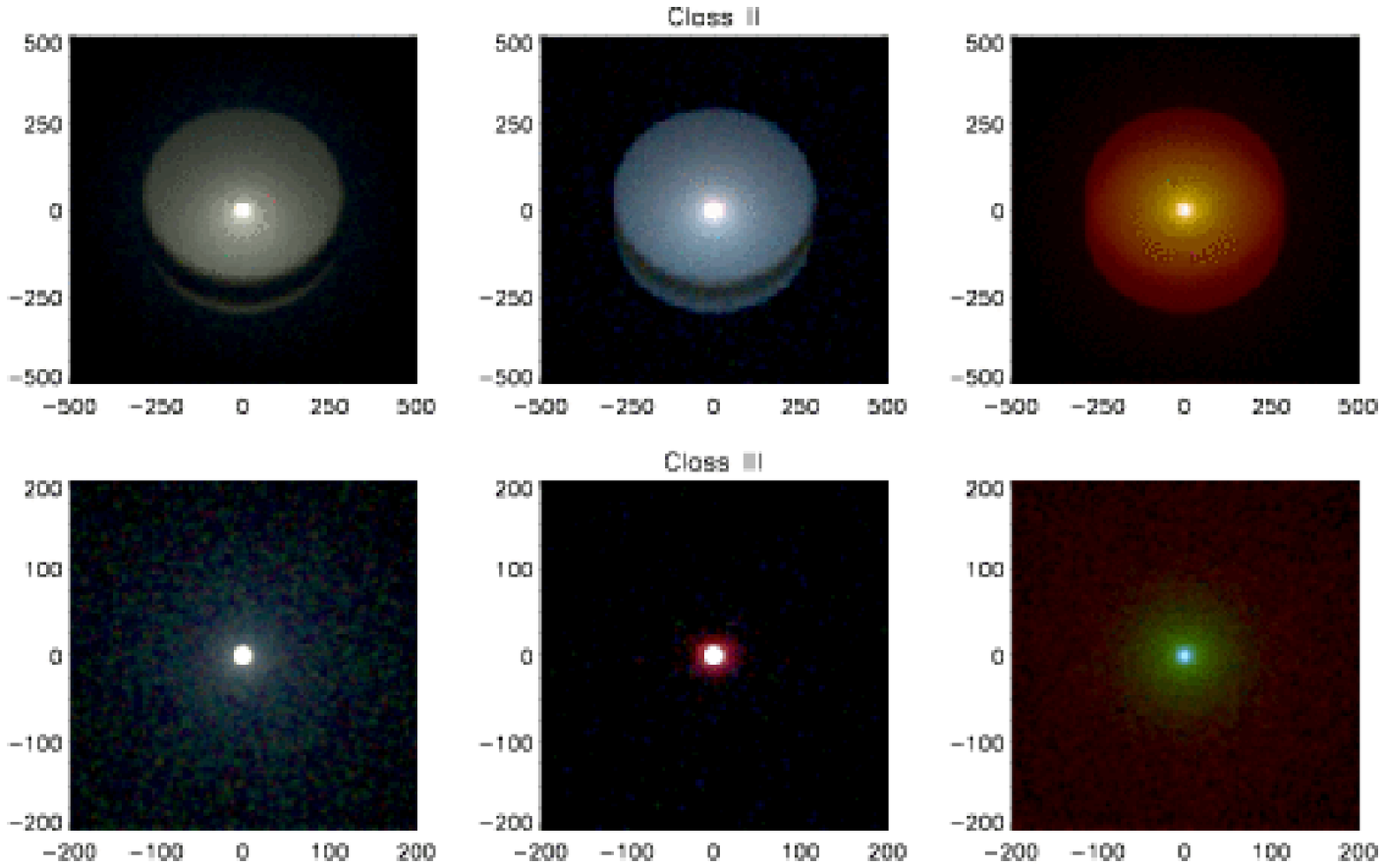}
\caption{}
\end{figure}
\end{document}